\documentclass[11pt,a4paper]{article}
\pdfoutput=1

\usepackage{jheppub}
\usepackage{latexsym}
\usepackage{multirow}
\usepackage{color}
\usepackage[usenames,dvipsnames,svgnames,table]{xcolor}
\usepackage{lipsum}
\usepackage{footnote}
\usepackage{graphicx}
\usepackage{epsfig}
\usepackage{epsf}
\usepackage{dcolumn}
\usepackage{bm}
\usepackage{textcomp}
\usepackage{float}
\usepackage{subfig}
\usepackage{lineno}
\usepackage{hypcap}
\usepackage{cancel}
\usepackage[]{hyperref}

\usepackage[T1]{fontenc}
\usepackage[normalem]{ulem}

\hypersetup{
  bookmarks=true,
  unicode=false,
  pdftoolbar=true,
  pdfmenubar=true,
  pdffitwindow=true,
  pdfstartview={FitH},
  pdfsubject={Neutrino Oscillations Phenomenology},
  pdfnewwindow=true,
  pdfcreator={RevTeX},
  colorlinks=true,
  linkcolor=blue,
  citecolor=red,
  filecolor=black,
  urlcolor=blue,
}
\usepackage{mathtools}
\usepackage{cancel}
\usepackage[utf8]{inputenc}

\def\lsim{\raise0.3ex\hbox{$\;<$\kern-0.75em\raise-1.1ex
\hbox{$\sim\;$}}}
\def\gsim{\raise0.3ex\hbox{$\;>$\kern-0.75em\raise-1.1ex
\hbox{$\sim\;$}}}

\newcommand{\beq}{\begin{equation}}
\newcommand{\eeq}{\end{equation}}
\newcommand{\beqa}{\begin{eqnarray}}
\newcommand{\eeqa}{\end{eqnarray}}

\title{Constraining visible neutrino decay at KamLAND and JUNO }

\author[a,b]{Yago P. Porto-Silva,}
\author[a]{Suprabh Prakash,}
\author[a]{O. L. G. Peres,}
\author[c]{Hiroshi Nunokawa,}
\author[d]{Hisakazu Minakata}
\affiliation[a]{Instituto de F{\'i}sica Gleb Wataghin - UNICAMP, 13083-859,
Campinas, S\~ao Paulo, Brazil}
\affiliation[b]{Max-Planck-Institut f{\"u}r Kernphysik, Saupfercheckweg 1, 69117 Heidelberg, Germany}
\affiliation[c]{Departmento de F\'isica,
Pontif\'icia Universidade Cat\'olica do Rio de Janeiro,
C.P. 38071, 22452-970, Rio de Janeiro, Brazil}
\affiliation[d]{
Center for Neutrino Physics, Department of Physics, Virginia Tech, Blacksburg, Virginia 24061, USA}

\emailAdd{yporto@ifi.unicamp.br}
\emailAdd{sprakash@ifi.unicamp.br}
\emailAdd{orlando@ifi.unicamp.br, ORCID:0000-0003-2104-8460}
\emailAdd{nunokawa@puc-rio.br}
\emailAdd{minakata71@vt.edu}

\abstract{
We study visible neutrino decay at the reactor neutrino experiments KamLAND and, JUNO. Assuming the Majoron model of neutrino decay, we obtain constraints on the couplings between Majoron and neutrino as well as on the lifetime/mass of the most massive neutrino state i.e., $\tau_{3} / m_{3}$ or $\tau_{2} / m_{2}$, respectively, for the normal or the inverted mass orderings.
We obtain the constraints on the lifetime $\tau_{2} / m_{2} \geq 1.4 \times 10^{-9}~\rm{s/eV}$ in the inverted mass ordering for both KamLAND and JUNO at 90\% CL. In the normal ordering in which the bound can be obtained for JUNO only, the constraint is milder than the inverted ordering case, $\tau_{3} / m_{3} \geq 1.0 \times 10^{-10}$~s/eV at 90\% CL.
We find that the dependence of lightest neutrino mass ($=m_{\rm{lightest}}$), $m_1 (m_3)$ for the normal (inverted) mass ordering, on the constraints for the different types of couplings (scalar or pseudo-scalar) is rather strong, but the $m_{\rm{lightest}}$ dependence on the lifetime/mass bound is only modest.
}

\arxivnumber{}

\begin{document}
\maketitle
\flushbottom
\newpage


\section{Introduction}
\label{Sec:intro}

In the Standard Model (SM) of particle physics, neutrinos are stable particles. This is not only true in the original formulation of the SM, in which neutrinos are massless but also true in practice in the neutrino mass embedded version, the $\nu$SM. In the latter, the neutrino has a finite lifetime due to a nonzero mass and the lepton flavor mixing. But, the lifetime is extremely long, $>10^{45}$s for radiative decay \cite{Petcov:1976ff,Marciano:1977wx,PhysRevD.16.1444,Shrock:1982sc}. Since such a very long lifetime is practically unmeasurable, neutrinos can be regarded as stable particles in the $\nu$SM. Therefore, if neutrino decay is detected it will imply evidence for new physics beyond the SM.

One can impose rather severe constraints on neutrino lifetime by
observations of astrophysical neutrinos from various distant sources, in
particular, SN1987A
\cite{Frieman:1987as,Hirata:1987hu,Bionta:1987qt,Berezhiani:1989za,Kachelriess:2000qc},
supernova in
general~\cite{Tomas:2001dh,Lindner:2001th,Ando:2003ie,Ando:2004qe,Fogli:2004gy,deGouvea:2019goq} and the sun \cite{Bahcall:1972my,Raghavan:1987uh,Berezhiani:1991vk,Joshipura:1992vn,Acker:1992eh,Berezhiani:1992ry,Berezhiani:1993iy,Choubey:2000an,Bandyopadhyay:2001ct,Beacom:2002cb,Joshipura:2002fb,Bandyopadhyay:2002qg,Das:2010sd,Berryman:2014qha,Picoreti:2015ika,Aharmim:2018fme,Funcke:2019grs}.
However, in this method, the lifetime bounds can be obtained only for $\nu_{2}$ ($\overline{\nu}_{2}$) or $\nu_{1}$ ($\overline{\nu}_{1}$), because they have a large component of $\nu_e$ ($\overline{\nu}_e$). It does not appear to be possible to obtain a robust bound on $\nu_{3}$ ($\overline{\nu}_{3}$) lifetime, which implies a serious limitation in the case of normal mass ordering (NO), $m_{3} >  m_{2} > m_{1}$. In this case, it is worthwhile to look for ways by which $\nu_{3}$ lifetime can be experimentally constrained. In fact, there have been many discussions and various methods are proposed to constrain $\nu_{3}$ lifetime, e.g., by using the astrophysical \cite{Hannestad:2005ex,Baerwald:2012kc,Dorame:2013lka,Bustamante:2016ciw,Pagliaroli:2015rca,Escudero:2019gfk}, atmospheric \cite{Barger:1998xk,Fogli:1999qt,Meloni:2006gv,Maltoni:2008jr,GonzalezGarcia:2008ru,Choubey:2017eyg,Denton:2018aml,Choubey:2018kah}, accelerator \cite{Gomes:2014yua,Phdabner,Pagliaroli:2016zab,Gago:2017zzy,Choubey:2017dyu,Choubey:2018cfz,deSalas:2018kri,Tang:2018rer}, and the reactor neutrinos \cite{Abrahao:2015rba}. In the case of inverted mass ordering (IO), $m_{2} >  m_{1} > m_{3}$, generally speaking, the astrophysical constraints on the lifetime of high mass states are powerful as stated above.

It appears that most of the foregoing analyses of $\nu_{3}$ lifetime were done under the assumption of invisible decay, namely, the case that decay products are unobservable. See, however, refs.~\cite{Gago:2017zzy,Coloma:2017zpg,Ascencio-Sosa:2018lbk,Huang:2018nxj,Funcke:2019grs} for the analyses with visible neutrino decay. Moreover, the majority of the works devoted to the analyses of neutrino decay so far restrict themselves to the case of NO.

In this paper, we discuss the bound on neutrino lifetime with visible
neutrino decay. We consider both mass orderings, NO and IO. To treat
visible neutrino decay we must specify the model which allows neutrinos to 
decay, and we use the Majoron model
\cite{Chikashige:1980qk,Chikashige:1980ui,Schechter:1981cv,Gelmini:1980re,Gelmini:1983ea,1988SvA....32..127D,Berezhiani:1990sy,Dias:2005jm}
as a concrete model of visible neutrino decay (see
Section~\ref{Sec:what-new}). To place the bound on neutrino decay
lifetime, we analyze the reactor neutrino experiments,
KamLAND~\cite{Gando:2013nba} and JUNO~\cite{An:2015jdp}. For the former
we use the real data in ref.~\cite{Gando:2013nba}, and for the latter
the simulated one assuming the total number of 140,000 events
which would be obtained with an exposure of 220 GW$\cdot$years
or somewhat more depending on the actual availability (which is expected 
to be $\sim$ 85-90 \%) of reactors.\footnote{
Because of this feature and for a very simplified code, our analysis may be called more properly as the one for the ``JUNO-like'' setting.}

Under the visible neutrino decay hypothesis, there appear a few new features in the analysis:
\begin{itemize}
\item
Unlike the case of invisible decay, the decay products include active neutrino states, which we call the ``daughter'' neutrinos,\footnote{
For notations of the ``parent'' and ``daughter'' neutrinos, see Section~\ref{Sec:general-formula} for the definitions. }
and they can produce additional events in the detectors;

\item
There is a clear difference in the constraints we will obtain between
     the cases of NO and IO. In the IO, $\overline{\nu}_{1}$ and
     $\overline{\nu}_{2}$ decay into $\overline{\nu}_{3}$ and ${\nu}_{3}$
, which leads to a significant deficit of inverse beta decay events due to the large $\overline{\nu}_{e}$ component in the parent $\overline{\nu}_{1}$ and $\overline{\nu}_{2}$ mass eigenstates.
Whereas in the NO, $\overline{\nu}_{3}$ decays into
     $\overline{\nu}_{1}$ and $\overline{\nu}_{2}$
as well as $\nu_1$ and $\nu_2$.
Since the parent $\overline{\nu}_{3}$ states are much less populated by $\overline{\nu}_{e}$ due to the small value of $\theta_{13}$, the effect of decay on the $\overline{\nu}_{e}$ spectrum is only modest.

\end{itemize}

Now, we must spell out our attitude on the astrophysical neutrino bound on neutrino decay. Though the bound is likely to be correct and is probably robust we do not use the lifetime bound as granted in our analysis. The reasons for doing this is twofold:
(1) The lifetime bound from the reactor neutrino experiments is completely independent of  the bounds obtained by the solar and the supernova data.
(2) The analysis to derive the solar neutrino bounds on the Majoron couplings is not simple. Most notably, the antineutrino appearance from the sun is involved, which requires a separate analysis. In a variety of contexts, it does make sense to obtain the laboratory bounds even though the astrophysical bounds are much stronger than the laboratory ones.\footnote{
If we consider the lifetime bound from the astrophysical neutrinos, the order of magnitude bound we would obtain in the relevant channel would be $\tau / m \gsim 10^{-4}$ s$/$eV and $\tau / m \gsim 10^{6}$ s$/$eV, for the solar and the supernova neutrinos, respectively. Such bounds are several orders of magnitude stronger than the laboratory bounds summarized in Table \ref{tab:bounds}.
}

In our analysis, for simplicity, we turn on only the Majoron couplings $g^{23}$ and $g^{13}$ (See Eq.(\ref{eq0}) for their definitions). In principle we can turn on all the couplings including $g^{12}$, but the analysis becomes far more complicated. It is also very likely that the qualitative features of the bound obtained for the Majoron couplings remain unchanged in our reduced setting.
Therefore, only the following decay modes are allowed in our setting:
$\overline{\nu}_{3} \rightarrow \overline{\nu}_{1}/\nu_{1} + \phi$ or
$\overline{\nu}_{3} \rightarrow \overline{\nu}_{2}/\nu_{2} + \phi$ in the NO, and
$\overline{\nu}_{1} \rightarrow \overline{\nu}_{3}/\nu_{3} + \phi$ or
$\overline{\nu}_{2} \rightarrow \overline{\nu}_{3}/\nu_{3} + \phi$ in the IO, where $\phi$ denotes a Majoron particle.

In this paper, after understanding all the above points, we concentrate on deriving the reactor neutrino bound on the Majoron couplings $g^{23}$ and $g^{13}$, and the corresponding $\tau/m$ in both the NO and the IO. Thus, we explore systematically for the first time, assuming visible neutrino decay, the constraints that can be imposed on $\overline{\nu}_{3}$ lifetime (in the case of NO) and on $\overline{\nu}_{2}$ and $\overline{\nu}_{1}$ lifetimes (in the case of IO) by using the medium- and long-baseline reactor anti-neutrinos experiments. Yet, we must mention that our analysis is based on the Majoron model, and is done under the assumption of switching off the coupling between $\nu_{1}$, $\nu_{2}$, and Majoron.

\section{Brief recollection of the existing bounds on neutrino decay }
\label{Sec:existing-bounds}

In most of the existing literatures, the bounds on neutrino decay have been calculated for NO, and hence Table~\ref{tab:bounds} contains the bound for the NO which uses the $\nu_{3}/\overline{\nu}_{3}$ decay mode only. The tabulated bounds in Table~\ref{tab:bounds} span the region from a few $\times10^{-12}$ to a few $\times~10^{-10}$~/eV. These bounds, which utilize the artificial neutrino beams, are very loose compared with the solar neutrino bounds~\cite{Bandyopadhyay:2001ct,Beacom:2002cb,Bandyopadhyay:2002qg,Joshipura:2002fb,Das:2010sd,Berryman:2014qha,Picoreti:2015ika,Aharmim:2018fme,Funcke:2019grs}. The latter which is usually quoted as the one for $\nu_{2}$ is:  $\tau_{2} / m_{2} \gsim 7.02 \times  10^{-4}$~s/eV at 99\% C.L.~\cite{Picoreti:2015ika}.

\begin{table}[tbp!] 
\setlength{\tabcolsep}{0.5pt}
\centering
\renewcommand{\arraystretch}{1.6}
\begin{tabular}{ l c c l }
\hline \hline
Analysis & Daughter $\nu$ included & Lower Limit  (s/eV) \\
\hline
Atmospheric and long-baseline data~\cite{GonzalezGarcia:2008ru}   & No &  $ 2.9 \times 10^{-10}$ (90\% C.L)\\
MINOS and T2K data~\cite{Gomes:2014yua} &   No & $ 2.8 \times 10^{-12} \quad \textrm{(90\%~C.L.)}$ \\
MINOS and T2K data~\cite{Gago:2017zzy} &   Yes  &  $1.5 \times 10^{-11}\quad \textrm{(90\%~C.L.)}$ \\
JUNO  expected sensitivity~\cite{Abrahao:2015rba}   &  No & $\ 7.5 \times 10^{-11}$ (95\% C.L.) \\
DUNE expected sensitivity~\cite{Coloma:2017zpg} &   Yes &  $ \left(1.95 - 2.6\right) \times 10^{-10}$ (90\% C.L.) \\
ICAL expected sensitivity~\cite{Choubey:2018kah} & No & $ 1.6 \times 10^{-10}$ (90\% C.L)\\
\hline
\hline
\end{tabular}
\caption{\label{tab:bounds}
\footnotesize{Current and prospective constraints (expected sensitivities)
on neutrino lifetime from
neutrino oscillation experiments.  The lowest (highest)
value for DUNE sensitivity is for the highest (lowest)  $m_1$ lightest
neutrino mass. All results assume the NO.}}
\end{table}

\section{Phenomenological aspects of visible neutrino decay }
\label{Sec:what-new}

To describe visible neutrino decay we use the Majoron model with
the following interaction Lagrangian
\begin{equation}
\mathcal{L}_{\rm{int}}= \left(\dfrac{g^{ij}_{\rm S}}{2}\right)\overline{\nu}_{i}\nu_{j}\phi +
\left(\dfrac{g^{ij}_{\rm PS}}{2}\right)\overline{\nu}_{i}i\gamma_{5}\nu_{j}\phi ,
\label{eq0}
\end{equation}
where $\phi$ is a Majoron field, and $g_{\rm S}^{ij}$ and $g_{\rm PS}^{ij}$ represent, respectively, scalar and pseudo-scalar couplings which are complex in general, with the neutrino mass eigenstate indices $i,j = 1,2,3$.
Given the model Lagrangian (\ref{eq0}), we have the  two-body decay modes $\overline{\nu}_{3}~\rightarrow~\overline{\nu}_{i}/\nu_{i}~+~ \phi~(i=1,2)$ in the case of NO, and $\overline{\nu}_{i} \rightarrow \overline{\nu}_{3}/\nu_{3} + \phi~(i=1,2)$ in the case of IO. As we stated in Section~\ref{Sec:intro}, we switch off the decay mode $\overline{\nu}_{2} \rightarrow \overline{\nu}_{1}/\nu_{1} + \phi~(i=1,2)$ in the IO. In this work, we assume that the Majoron is  massless.

\begin{figure}[htbp!]
\vglue -0.5cm
\centering
\includegraphics[scale=1.0]{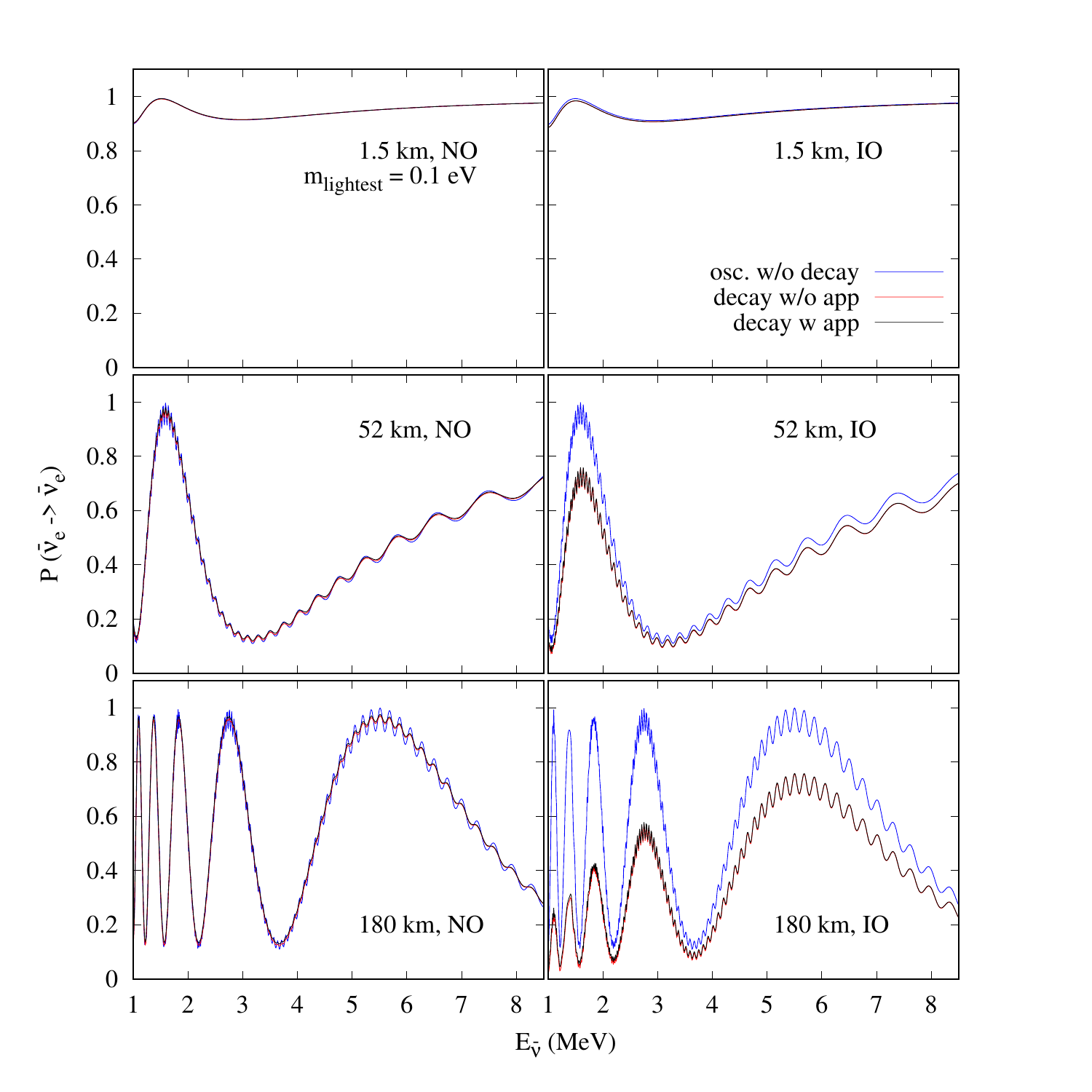}
\vglue -0.5cm
\caption{\footnotesize{The {\it effective} electron anti-neutrino disappearance probability
as a function of anti-neutrino energy for the Daya Bay (top panel),
JUNO (middle panel) and KamLAND (bottom panel) experiments.
The left (right) panel is for the NO (IO). We show the effective probabilities (defined as the ratio of the $\overline{\nu}_e$ flux arriving at the detector
divided by the original flux at the detector in the absence of oscillation for a given
neutrino energy) assuming standard oscillations without any decay
(labeled ``osc. w/o decay'' in blue), with visible decay effect but
without the daughter neutrino contribution (labeled ``decay w/o app'' in red)
and with visible decay including the daughter neutrino contribution
(labeled ``decay w app'' in black).
For the visible decay, we consider $g_{\rm S} = g_{\rm PS} = 0.2$ for
NO and $g_{\rm S} = g_{\rm PS} = 0.1$ for IO.
Note that in some of the plots, the individual curves are too close
together to be distinguished.}}
\label{fig:prob}
\end{figure}

Phenomenology of neutrino decay depends crucially on the following two factors,

\begin{itemize}
\item
if neutrinos undergo visible or invisible decay, that is  if the decay products are experimentally detectable or not\footnote{
When neutrinos undergo visible decay, it is sometimes argued that even active daughter neutrino may be unobservable when its energy is too low to be detected. However, the terminology of calling it as ``invisible decay'' may be confusing because observability depends on the experimental settings and/or detector performances. For this reason, we always classify neutrino decay into active neutrino species as ``visible decay'' for clarity.
},
\item
if the neutrino masses exhibit NO or IO.

\end{itemize}
\noindent
We emphasize that the above, seemingly-obvious statements do indeed provide the key to understand the results in this paper.
This fact is best summarized in Figure~\ref{fig:prob} in which the
$\overline{\nu}_{e}$ disappearance probabilities in the absence or
presence of the decay are plotted as a function of the anti-neutrino
energy at a few characteristic distances: The top ($L=1.5$ km), middle
($L=$ 52 km) and the bottom ($L=180$ km) panels
correspond, respectively, to the far detectors in Daya Bay, JUNO, and the KamLAND
experiments.
The left (right) panels in Figure~\ref{fig:prob} are for the NO (IO).
We note that the probability is shown in Figure~\ref{fig:prob} is the {\em
effective} one in the sense that it is defined as the ratio of the
$\overline{\nu}_e$ flux at
the detector with oscillation plus decay effects to the flux without
them. 
The former includes the contribution of daughter neutrinos which exists in the case of visible decay.

We first observe that the effect of decay is very noticeable in the case
of IO (right panels) despite that the assumed magnitude of couplings for
the IO case is smaller than that for NO
, as can be seen in the right panels of Figure~\ref{fig:prob}. It is because decay of the higher mass eigenstates $\overline{\nu}_{1}$ and $\overline{\nu}_{2}$, which occurs copiously in the reactor-produced $\overline{\nu}_{e}$ flux, leads to a much stronger reduction of the survival probability $P(\overline{\nu}_{e} \rightarrow \overline{\nu}_{e})$ than the NO case (see below). The effect can be seen clearly in the right panels of Figure~\ref{fig:prob} with the Majoron coupling constants $g_{\rm S}=g_{\rm PS}=0.1$.
In the case of NO (left panels), on the other hand, the effect of decay is small, irrespective of whether the contribution from the daughter neutrinos is included or not. Unlike the case of IO, the reduction of the $\overline{\nu}_{e}$ flux is minor as the parent $\overline{\nu}_{3}$ component is small in reactor $\overline{\nu}_{e}$ due to suppression by small $\vert U_{e3} \vert^2 = \sin^2\theta_{13}$. Therefore, the effect of visible decay in the case of NO is just to dampen the atmospheric-$\Delta m^2_{31}$ driven neutrino oscillation~\cite{Abrahao:2015rba}.

In visible decay, an additional effect, a pile-up of events at low energies due to the daughter neutrino contribution, should be observed\footnote{
We remark here that in the case of reactor neutrino experiments
for the baseline of $\sim O(100)$ km,
the area covered by the decay beam spread is expected to be
$\lsim O (10^{-2})~m^2$ which is much smaller than the detector
sizes of KamLAND/JUNO.
Therefore, we assume that in practice all the daughter neutrinos 
which are produced from the parent neutrinos emitted from the source toward the direction of the detector,
reach the detector \cite{Lindner:2001fx}.}.
For the IO, for our choice of couplings $g_{\rm{S}}=g_{\rm{PS}}=0.1$
this effect is barely noticeable by eyes in Figure~\ref{fig:prob}
but only for longer baseline as in the case of KamLAND at lower
energies as a small difference between the cases without (red curves)
and with (black curves) daughter contributions.
We confirmed that by using somewhat larger values of couplings 
the pile-up effect become more prominent but its effect is tiny 
in any case because of small ($\propto \vert U_{e3} \vert^2$) $\overline{\nu}_{e}$ component in the decay product $\overline{\nu}_{3}$. The effect is negligible for the NO because the decay effect itself is small. 
However, we include the daughter neutrino contribution for both NO as well as IO, irrespective of its importance -- on which some comments will follow later. 


\section{The oscillation probabilities with neutrino decay}
\label{Sec:probabilities}

We first recapitulate the formulas of the neutrino oscillation probabilities in the simultaneous presence of flavor oscillations and decay~\cite{Kim:1990km,Lindner:2001fx,PalomaresRuiz:2005vf,Coloma:2017zpg}. We treat the system as in vacuum which is a good approximation for the reactor neutrino experiments\footnote{While we have ignored the Earth matter effects to obtain the results shown in this work, we have checked explicitly their impact on decay
by following \cite{Ascencio-Sosa:2018lbk} and found that the decay rates change very little (much less than 1\%) due to the matter effects.
We verified that the modification of electron anti-neutrino survival probabilities due to the matter effects are typically less than 1\% in
the relevant energy range for both the standard situation as well as in the presence of decay for the experiments we considered in this work.}.

\subsection{Neutrino decay: General formula}
\label{Sec:general-formula}

We start by examining a generic case in which each of the three massive
neutrinos oscillate and decay at the same time. 
When a neutrino of flavour $\alpha$ with energy $E_{\alpha}$ is produced at the distance $L=0$, the differential probability that a neutrino of flavour $\beta$ with energy in the interval $E_{\beta}+dE_{\beta}$ is detected at the distance $L$, can be written as~\cite{Lindner:2001fx,PalomaresRuiz:2005vf,Coloma:2017zpg}
\begin{eqnarray}
\nonumber
\frac{dP_{\nu_{\alpha}^{r}\rightarrow\nu_{\beta}^{s}}}{dE_{\beta}} \left(E_{\alpha}, E_{\beta}, L \right) &=&
\left|\sum_{i}U_{\beta i}^{(s)}U_{\alpha i}^{(r)*}\exp\left[-i\frac{m_i^2L}{2E_{\alpha}}\right]
\exp\left[-\frac{1}{2}\left(\frac{\tau_{i}}{m_{i}}\right)^{-1}\frac{L}{E_{\alpha}}\right]\right|^2
\delta(E_{\alpha} - E_{\beta})~\delta_{rs} \\
&+&
\int^L_0 dL^{\prime} \left|\mathcal{A}_{\nu_{\alpha}^{r}\rightarrow\nu_{\beta}^{s}}\left(E_{\alpha}, E_{\beta}, L^{\prime}\right)\right|^2.
\label{degenral-P-formula}
\end{eqnarray}
In Eq.~(\ref{degenral-P-formula}), $\tau_{i}$ and $m_{i}$ represent $\nu_{i}$'s proper lifetime and mass, respectively, and the indices $r$ and $s$ specify, respectively, parent and daughter neutrino helicities. The matrix element $U_{\beta i}^{(s)}=U_{\beta i} (U_{\beta i}^\ast)$ corresponds to the case for positive (negative) helicity.
The first term of the differential probability in Eq.~(\ref{degenral-P-formula}) describes the contribution from a parent neutrino of flavor $\alpha$ which survived after propagating a distance $L$.
%
%
The second term in Eq.~(\ref{degenral-P-formula}) is the daughter contribution and contains the decay amplitude $\mathcal{A}_{\nu_{\alpha}^{r}\rightarrow\nu_{\beta}^{s}}\left(E_{\alpha},
E_{\beta}, L^{\prime}\right)$ defined by
\begin{equation}
  \mathcal{A}_{\nu_{\alpha}^{r}\rightarrow\nu_{\beta}^{s}}\left(E_{\alpha}, E_{\beta}, L^{\prime}\right)  = \underset{i\neq j}{\sum^{3}_{i=1}\sum^{3}_{j=1}} \left(U^{r}_{\alpha i}\right)^{\ast}\left(U^{s}_{\beta j}\right)e^{-iE^s_{j}\left(L-L^{\prime}\right)}e^{-\frac{\alpha^s_{j}(L-L^{\prime})}{2E^s_{j}}}\sqrt{\frac{\alpha^{rs}_{ij}}{E^r_{i}}}\sqrt{\eta_{ij}}e^{-iE^r_{i}L^{\prime}}e^{-\frac{\alpha^r_{i}L^{\prime}}{2E^r_{i}}}.
\label{visible-amp}
\end{equation}
It describes contribution of daughter neutrino of energy $E_{\beta}$ produced by decay of a parent neutrino with energy $E_{\alpha}$ at a distance $L' (< L)$.
Here, $\alpha/E$ represents the partial or full decay rates and $\eta$ represents normalized energy distribution of the daughter neutrinos. To understand Eq. \ref{visible-amp} in detail, please refer to Ref. \cite{Lindner:2001fx}.
The first term of Eq.~(\ref{degenral-P-formula}) is often called the ``invisible'' contribution. But, the case that we examine in this paper has no invisible decay; the decay products always include active neutrinos and hence are always visible in principle.
To prevent confusion we call the first and the second terms of Eq.~(\ref{degenral-P-formula}) as the ``parent'' and ``daughter'' contributions, respectively.
We remark that if a neutrino undergoes invisible decay, the differential probability is given by the first term of Eq.~(\ref{degenral-P-formula}).
Then, what is the difference between the invisible decay and the parent contribution of our visible decay?
The answer is that in the case of un-observable final states (such as sterile neutrinos), the decay width $\Gamma = 1/\tau$ does not contain information about the final states. Whereas in our case $\Gamma$, which is computed with the Majoron model, does contain information of final states, such as the mass of the daughter neutrino. The lightest neutrino mass dependence of the event spectrum will be demonstrated in Section~\ref{Sec:results}.

\subsection{Parent contribution in visible neutrino decay }

Let us calculate the contribution from the parent neutrinos i.e, the first term in Eq.~(\ref{degenral-P-formula}).
It gives the whole contribution in the case of invisible neutrino decay.
By the nature of this term, helicity flip cannot be involved in it. Then, after integration over the neutrino energy $E_{\alpha}$ we obtain ($i,j=1,2,3$)

\begin{eqnarray}
P_{\alpha\beta}^{ \text{parent} } &=&
\left|\sum_{i} U_{\beta i} U_{\alpha i}^* \exp\left[-i\frac{m_i^2L}{2E_{\alpha}}\right]\exp\left[-\frac{1}{2}
\left(\frac{\tau_{i}}{m_{i}}\right)^{-1}\frac{L}{E_{\alpha}}\right]\right|^2
\nonumber \\
&=&
\nonumber \sum_{i} \vert U_{\alpha i} \vert^2 \vert U_{\beta i} \vert^2
\exp\left(-\frac{m_{i}}{\tau_{i}}\frac{L}{E_\alpha}\right) \\
&& + 2 \sum_{j > i}
U_{\beta i} U_{\alpha i}^* U_{\beta j}^* U_{\alpha j}
\exp\left\{-\left(\frac{m_{j}}{\tau_{j}}+\frac{m_{i}}{\tau_{i}}\right)
\frac{L}{2E_\alpha}\right\}\cos\left(\frac{\Delta m^2_{ji}L}{2E_\alpha}\right).
\label{parent-contribution}
\end{eqnarray}

Using the standard parameterization of the flavor mixing matrix \cite{Tanabashi:2018oca}, and substituting $\tau_{1},\tau_{2} \rightarrow \infty$ for the NO, we obtain for $\overline{\nu}_{e} \rightarrow \overline{\nu}_{e}$ channel\footnote{
The survival probabilities for neutrinos and anti-neutrinos are equal in vacuum due to CPT symmetry.}
\begin{eqnarray}
P_{\overline{e}\overline{e}}^{ \text{parent}}\text{ (NO) } =
&& \cos^4\theta_{12}\cos^4\theta_{13} + \sin^4\theta_{12}\cos^4\theta_{13}
+ \sin^4\theta_{13}\exp\left(-\frac{m_{3}}{\tau_{3}}\frac{L}{E_\alpha}\right)
\nonumber \\
&+& \dfrac{1}{2}\sin^22\theta_{12}\cos^4\theta_{13}
\cos\left(\frac{\Delta m^2_{21}L}{2E_\alpha}\right)
\nonumber \\
&+& \dfrac{1}{2}\sin^22\theta_{13}\cos^2\theta_{12}
\exp\left\{-\left(\dfrac{m_{3}}{\tau_{3}}\right)\frac{L}{2E_\alpha}\right\}
 \cos\left(\frac{\Delta m^2_{31}L}{2E_\alpha}\right)
\nonumber \\
&+& \dfrac{1}{2}\sin^22\theta_{13}\sin^2\theta_{12}
\exp\left\{-\left(\frac{m_{3}}{\tau_{3}}\right)\frac{L}{2E_\alpha}\right\}
\cos\left(\dfrac{\Delta m^2_{32}L}{2E_\alpha}\right).
\label{eq10}
\end{eqnarray}
And for the inverted mass ordering,
on substituting $\tau_{3} \rightarrow \infty$, we get
\begin{eqnarray}
P_{\overline{e}\overline{e}}^{\rm{parent} } \text{ (IO)} =
&&
\sin^4\theta_{13} \nonumber \\
&+& \cos^4\theta_{13}\left[\cos^4\theta_{12}
\exp\left(-\frac{m_{1}}{\tau_{1}}\frac{L}{E_\alpha}\right)
+
\sin^4\theta_{12}\exp\left(-\frac{m_{2}}{\tau_{2}}\frac{L}{E_\alpha}\right)\right]
\nonumber \\
&+&
\frac{1}{2}\sin^22\theta_{12}\cos^4\theta_{13}\exp\left\{-\left(\frac{m_{1}}{\tau_{1}}+
\frac{m_{2}}{\tau_{2}}\right)\frac{L}{2E_\alpha}\right\}\cos\left(\frac{\Delta m^2_{21}L}{2E_\alpha}\right)
\nonumber \\
&+&
\frac{1}{2}\sin^22\theta_{13} \cos^2\theta_{12} \exp\left\{-\left(\frac{m_{1}}{\tau_{1}}\right)\frac{L}{2E_\alpha}\right\}
\cos\left(\frac{\Delta m^2_{31}L}{2E_\alpha}\right)
\nonumber \\
&+&
\frac{1}{2}\sin^22\theta_{13}\sin^2\theta_{12}\exp\left\{-\left(\frac{m_{2}}
{\tau_{2}}\right)\frac{L}{2  E_\alpha}\right\}\cos\left(\frac{\Delta m^2_{32}L}{2E_\alpha}\right).
\label{parent-contribution-NO-IO}
\end{eqnarray}

\subsection{Daughter contribution in visible neutrino decay}

We calculate the contribution of daughter neutrinos, the second term in Eq.~(\ref{degenral-P-formula}). We assume that the couplings between the neutrinos and the Majoron are real quantities and hence there are no decay-related complex phases. Since we are interested only in the electron antineutrino disappearance probabilities here we drop the helicity indices $r$ and $s$ keeping in mind that the quantities correspond to antineutrinos and that only helicity preserving decays can be observed in reactor experiments. However, it should be noted that the full decay-widths include the sum over helicity-preserving as well as helicity-flipping partial decay-widths. In this work, we assume the CP-violating phase $\delta_{CP}$ as well as the Majorana phases to be 0. Thus, we can also ignore the complex conjugation of the flavor matrix elements.

The transition amplitude $\mathcal{A}_{\nu_{\alpha}\rightarrow\nu_{\beta}}\left(E_{\alpha}, E_{\beta},L^{\prime}\right)$ for the NO where $\overline{\nu}_{3}$ decays to $\overline{\nu}_{1}$ or $\overline{\nu}_{2}$, is given by~\cite{Lindner:2001fx,PalomaresRuiz:2005vf,Coloma:2017zpg}
\begin{equation}
\mathcal{A}_{ \nu_{\alpha}\rightarrow\nu_{\beta}}^{\rm NO}\left(E_{\alpha}, E_{\beta},L^{\prime}\right)
=\sum_{d=1,2}
 U_{\alpha 3}U_{\beta d} \sqrt{\Gamma_{3 d}}\sqrt{W_{3 d}}~e^{-iE_{\beta}\left(L-L^{\prime}\right)}
e^{-iE_{\alpha}L^\prime}e^{-\Gamma_{3}L^\prime/2}.
\label{amplitude-normal}
\end{equation}
%
Whereas the transition amplitude for the IO, where $\overline{\nu}_{2}$ or $\overline{\nu}_{1}$ decay to $\overline{\nu}_{3}$, takes the form
\begin{equation}
\mathcal{A}_{ \nu_{\alpha}\rightarrow\nu_{\beta}}^{\rm IO}\left(E_{\alpha}, E_{\beta},L^{\prime}\right)
=\sum_{p=1,2}U_{\alpha p}U_{\beta 3} \sqrt{\Gamma_{p 3}}\sqrt{W_{p 3}}~e^{-iE_{\beta}\left(L-L^{\prime}\right)}e^{-iE_{\alpha}L^\prime}e^{-\Gamma_{p}L^\prime/2}.
\label{amplitude-inverted}
\end{equation}
Here $E_{\alpha}$ $=p_{\alpha}+m_p^2/(2p_{\alpha})$ represents the energy of the parent neutrinos while $E_{\beta}$  $=p_{\beta}+m_d^2/(2p_{\beta})$ represents the energy of the daughter neutrinos. $p_{\alpha}$ ($p_{\beta}$) represent the amplitude of the three-momentum of the parent (daughter) neutrinos while $m_{p}$ ($m_{d}$) represent their constituent mass eigenvalues respectively. We assume that the different mass eigenstates possess the same momentum and they are relativistic; thus substituting $p$ for $E$.
In the above equations, $\Gamma_{ij}$ is the partial decay width
and $W_{ij}$ represents the normalized energy distribution function
for the daughter neutrino for the decay
$\overline{\nu}_{i}\rightarrow\overline{\nu}_{j}$. The explicit formulas are given in the Appendix~\ref{apa}.

Eqs.~(\ref{amplitude-normal}) and (\ref{amplitude-inverted}) describe the process in which $\nu_{\alpha}$ is produced at $L=0$, propagates as the parent neutrino $\nu_{p}$ to $L'$, then it decays into the daughter state $\nu_{d}$ at this distance $L'$ and is detected as $\nu_{\beta}$ at $L$ > $L'$ after traversing the distance $L - L^{\prime}$. In the NO, $p=3$ and $d=1,2$, while in the IO $p=1,2$ and $d=3$.

Using the Eq.~(\ref{amplitude-normal}) we can compute the visible decay term in Eq.~(\ref{degenral-P-formula}), we find that for the NO,
\begin{equation}
\int^L_0 \left|\mathcal{A}_{ \nu_{\alpha}\rightarrow\nu_{\beta}}\right|^2 dL^{\prime}
=
\left(U_{\alpha 3}\right)^2\sum^2_{j=1}\sum^2_{k=1} U_{\beta j} U_{\beta k}\sqrt{\Gamma_{3j}\Gamma_{3k}}\sqrt{W_{3j}W_{3k}}\left(\frac{e^{-\Gamma_{3}L}-e^{-i\frac{\Delta m^2_{jk}L}{2E_{\beta}}}}{i\frac{\Delta m^2_{jk}}{2E_{\beta}}-\Gamma_{3}}\right)
\label{eq5}
\end{equation}
and for the IO,
\begin{equation}
\int^L_0 \left|\mathcal{A}_{\nu_{\alpha}\rightarrow\nu_{\beta}}\right|^2 dL^{\prime}
=
\left(U_{\beta 3}\right)^2\sum^2_{j=1}\sum^2_{k=1}U_{\alpha j}U_{\alpha k}\sqrt{\Gamma_{j3}\Gamma_{k3}}\sqrt{W_{j3}W_{k3}}\left(\frac{1-e^{-\left(i\frac{\Delta m^2_{jk}}{2E_{\alpha}}+\frac{\Gamma_j+\Gamma_k}{2}\right)L}}{i\frac{\Delta m^2_{jk}}{2E_{\alpha}}+\frac{\Gamma_j+\Gamma_k}{2}}\right)
\label{eq6}
\end{equation}
In Eqs. (\ref{eq5}) and (\ref{eq6}) above, the imaginary terms change sign under an interchange of the indices $j$ and $k$; hence the sums are real.


\section{Sketchy descriptions of KamLAND and JUNO}
\label{experimentdetails}

In this section, we briefly describe the details of KamLAND and JUNO, which are phenomenologically relevant to our work.

\subsection{KamLAND}

The KamLAND (Kamioka Liquid Scintillator Antineutrino Detector) reactor
neutrino experiment consists of 1 kton of highly purified liquid
scintillator detector based in Japan. KamLAND detects neutrinos coming
from 16 nuclear power plants with a range of distances that go from
$140$ km to $215$ km. The average distance corresponds to $\sim 180$
km. 
The experiment ran in the reactor anti-neutrino mode from 2002 to 2012, 
collecting a total exposure of $4.90\times10^{32}$ target-proton-years. 
We consider the data presented in \cite{Gando:2013nba} to perform our
analysis of neutrino decay.
The information regarding backgrounds and systematic uncertainties have
also been taken from \cite{Gando:2013nba}.
The expected advantage of KamLAND over JUNO to study the decay
effect is, as we could see in the plot of probabilities in
Figure~\ref{fig:prob}, the longer average baseline, which is about 3.4
times the JUNO's baseline, leading to larger decay effects. 

\subsection{JUNO}

The JUNO (Jiangmen Underground Neutrino Observatory) experiment
\cite{An:2015jdp} is a future neutrino experiment that will be based in
China. It is expected to start taking data from the year 2022. JUNO has
been designed with the primary goal to measure the neutrino mass
ordering but it will also be able to measure the oscillation parameters
such as $\theta_{12}$, $\Delta m^2_{21}$ and $\Delta m^2_{31}$
with much better precision. The detector consists of a 20 kton fiducial mass of liquid scintillator and is located at an average distance of $\sim 53$ km from Yangjiang and Taishan nuclear power plants. The remote reactor cores at Daya Bay and Huizhou will also have a small contribution to the total flux arriving at the JUNO detector. In our experimental set-up, we consider the various reactor cores with different thermal powers and baselines as described in Table 2 of \cite{An:2015jdp}. The description of backgrounds and systematic uncertainties have been taken from \cite{An:2015jdp}. The exposure is set such that a total of $140,000$ events are obtained (including the backgrounds).

The signal in both of the experiments is the inverse beta-decay
(IBD) events in the energy range $\sim\left[1.8, 8\right]$ MeV, which essentially
determines the electron antineutrino disappearance probability for a
given, relatively well known IBD reaction cross-sections. The main
background in a search for visible neutrino decay in JUNO is the
geo-neutrino events at low energies.
We consider their contributions similarly as done
in \cite{Abrahao:2015rba}.
The expected advantage of JUNO over KamLAND to study the neutrino
decay is (i) much larger statistics and (ii) better energy resolution
which is crucial for the case of NO as we will see later.

To simulate the KamLAND and the JUNO experiments, we have used the GLoBES \cite{Huber:2004ka,Huber:2007ji} software package. The event rates and statistical-$\chi^2$ calculations have also been performed using GLoBES.

\section{Features of event rates in the presence of decay}
\label{eventsfeatures}

In this section, using the observed and the expected event rates at
KamLAND and JUNO, respectively, we stress upon the following
two features that crucially affect the results in the presence of visible decay.
\begin{enumerate}
\item
Dependence of event rates on the neutrino mass ordering, and
\item
Dependence of event rates on the lightest neutrino mass\footnote{The quantity that the experiment can constrain is $\tau/m$. It should be kept in mind that the dependence on the lightest neutrino mass is correlated with the chosen value of $g_{\rm S}$ and $g_{\rm PS}$.}.
\end{enumerate}
Note that the lightest neutrino is $\nu_{3}$ in the case of IO, and $\nu_{1}$ in the case of NO.

\begin{figure}[htbp!]
\vspace{-6mm}
\centering
\includegraphics[scale=0.7]{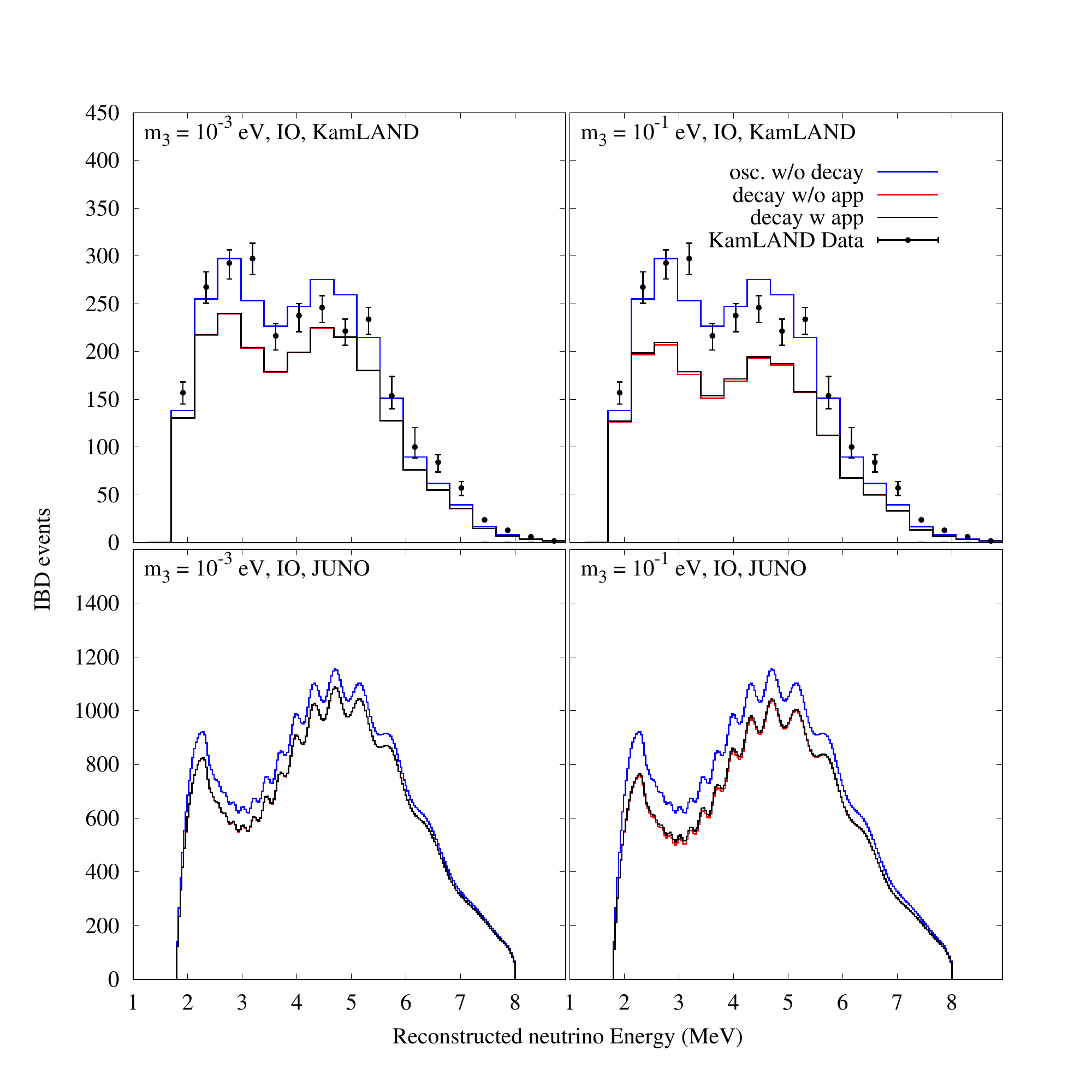}
\vspace{-6mm}
\caption{\footnotesize{Events rates vs. reconstructed neutrino energy assuming IO for KamLAND (top panel) and JUNO (bottom panel) with and without the decay
 effects. The left (right) panels correspond to the choice of  $m_{\rm
 lightest}=10^{-3}$~eV ($m_{\rm lightest}=10^{-1}$~eV). We show the rates for standard oscillations without any decay (labeled ``osc. w/o decay'' in blue), visible decay without the daughter neutrino contribution (labeled ``decay w/o app'' in red) and visible decay including the daughter contribution (labeled ``decay w app'' in black). For the visible decay, we consider $g_{\rm S} = g_{\rm PS} = 0.1$. 
 For these values of the couplings and $m_{\rm lightest}$, the values of $\tau_{2}/m_{2}$ are  $6.6\times10^{-10}\rm{s/eV}$ (for $m_{3} = 10^{-3}\rm{eV}$) and $3.7\times10^{-10}\rm{s/eV}$ (for $m_{3} = 10^{-1}\rm{eV}$).
 Also shown are the observed KamLAND data indicated by the black solid circles with error bars which are taken from \cite{Gando:2013nba}.}
}
\label{fig:events_io}
\end{figure}

\begin{figure}[htbp!]
\centering
\includegraphics[scale=0.7]{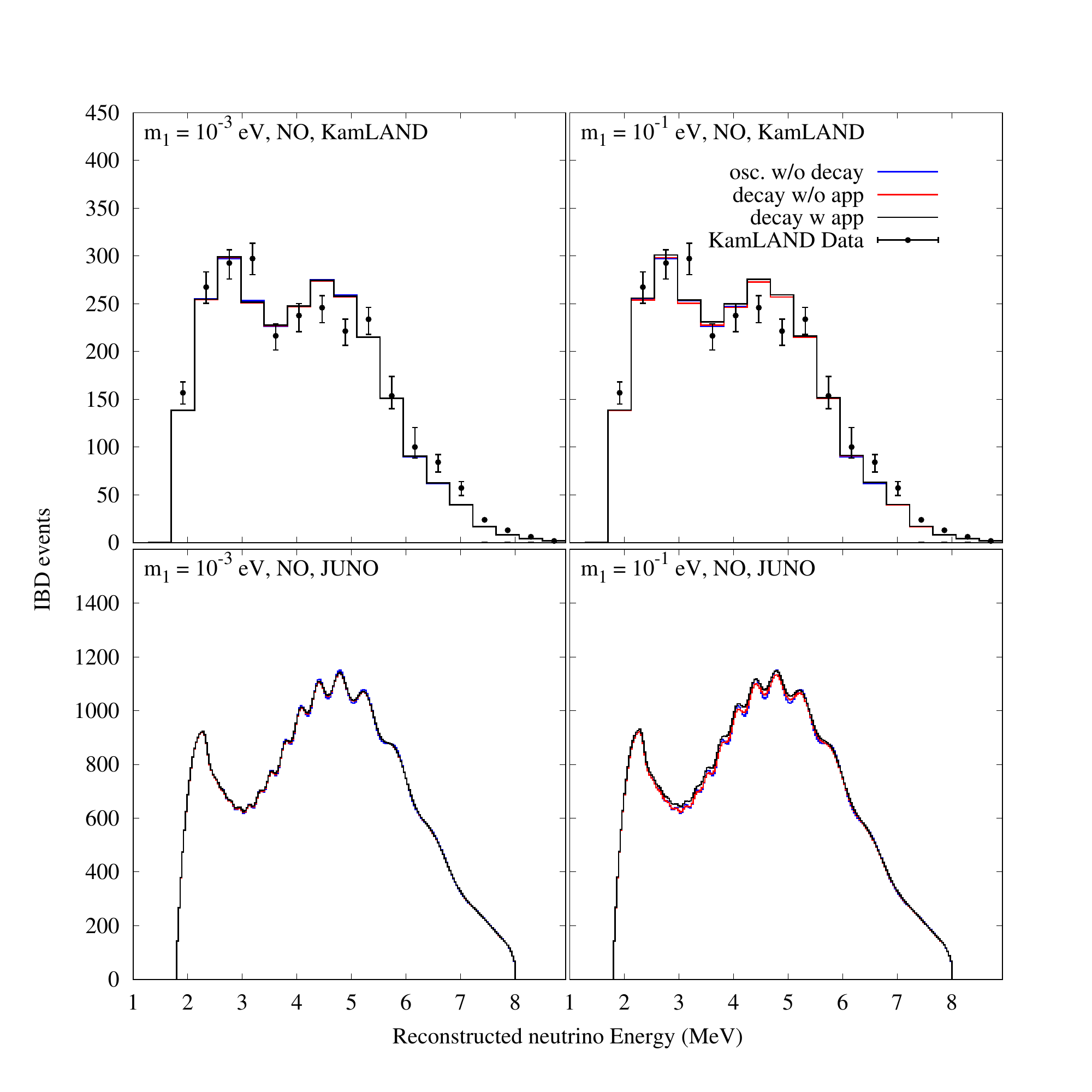}
\vspace{-6mm}
\caption{\footnotesize{Events rates vs. reconstructed neutrino energy assuming NO for KamLAND (top panel) and JUNO (bottom panel) with and without the decay
 effects. The left (right) panels correspond to the choice of  $m_{\rm
 lightest}=10^{-3}$~eV ($m_{\rm lightest}=10^{-1}$~eV). We show the rates for standard oscillations without any decay (labeled ``osc. w/o decay'' in blue), visible decay without the daughter neutrino contribution (labeled ``decay w/o app'' in red) and visible decay including the daughter contribution (labeled ``decay w app'' in black). For the visible decay, we consider $g_{\rm S} = g_{\rm PS} = 0.2$. 
 For these values of the couplings and $m_{\rm lightest}$, the values of $\tau_{3}/m_{3}$ are  $8.3\times10^{-11}\rm{s/eV}$ (for $m_{1} = 10^{-3}\rm{eV}$) and $4.6\times10^{-11}\rm{s/eV}$ (for $m_{1} = 10^{-1}\rm{eV}$).
 Also shown
 are the observed KamLAND data indicated by the black solid circles with
 error bars which are taken from \cite{Gando:2013nba}.
}}
\label{fig:events_no}
\end{figure}

In Figs.~\ref{fig:events_io} and ~\ref{fig:events_no}, we show the event rates for the IO and NO respectively. The top panels in both of these figures show the observed event rates at the KamLAND experiment with 17
bins of 425 MeV each lying in the reconstructed energy range $\left[1.7,
8.925\right]$ MeV. The bottom panels of these figures show the expected
event rates for the JUNO experiment with a total of 200 bins
(corresponding to the bin width of 0.031 MeV) in the reconstructed
energy interval $\left[1.8, 8.0\right]$ MeV. In Figs.~\ref{fig:events_io} and ~\ref{fig:events_no},
the event rates include the signal as well as the background
geo-neutrinos. In the left and the right panels, we assume\footnote{We remark here that the case of $m_{\rm{lightest}} =
10^{-3}$~eV closely mimics the results for $m_{\rm{lightest}} <
10^{-3}$~eV and hence aptly provides the lower limit consideration of
the neutrino masses \cite{Coloma:2017zpg}.}
$m_{\rm{lightest}} = 10^{-3}$~eV
and $m_{\rm{lightest}} = 10^{-1}$~eV, respectively. We assume the following values of the oscillation \cite{deSalas:2017kay} and decay parameters to generate these event rates.
\begin{itemize}
\item Inverted Ordering: $\theta_{12}=34.5^{\circ}$, $\theta_{13}=8.53^{\circ}$, $\theta_{23}=47.9^{\circ}$, $\Delta m^2_{21}=7.55~ \times~ 10^{-5}~\rm{eV}^2$, $\vert\Delta m^2_{31}\vert=2.42 \times 10^{-3}~\rm{eV}^2,~g_{\rm{S}}=g_{\rm{PS}}=0.1 $ .
\item Normal Ordering: $\theta_{12}=34.5^{\circ}$, $\theta_{13}=8.45^{\circ}$, $\theta_{23}=47.7^{\circ}$, $\Delta m^2_{21}=7.55~ \times~ 10^{-5}~\rm{eV}^2$, $\vert\Delta m^2_{31}\vert=2.50 \times 10^{-3}~\rm{eV}^2,~g_{\rm{S}}=g_{\rm{PS}}=0.2$.
\end{itemize}

From Figure~\ref{fig:events_io},
as expected from the probabilities shown in the right panels of Figure~\ref{fig:prob},
we see that the effects of decay are significant for IO. This is because, for IO, $\overline{\nu}_2$ or $\overline{\nu}_{1}$ mass eigenstate decays to $\overline{\nu}_3/\nu_{3}$. Thus, the decay is expected to affect the $\Delta m^2_{21}$-driven oscillations. For KamLAND and JUNO, these oscillations are much larger in magnitude compared to $\Delta m^2_{31}$-driven-oscillations due to the large value of $\theta_{12}$.
Therefore, decay effects are also large. Furthermore, when one considers
the full visible decay including the contribution from daughter
neutrinos, a pile-up of events at lower energies is noticeable for
$m_{\rm{lightest}}$ =$10^{-1}$ eV
(see  text below for the $m_{\rm{lightest}}$
dependence on the decay effect).
However, this is still a small effect as the appearance of $\overline{\nu}_{e}$ is suppressed due to the smallness of $\vert U_{e3}\vert^2$.
We note that the case considered to generate the results shown in
Figure~\ref{fig:events_io}, $g_{\rm{S}}=g_{\rm{PS}}=0.1$ for the IO case,
is turned out to be excluded as we will see later.

The decay effects in the event spectrum also depend on the lightest
neutrino mass as can be seen by comparing the left and right panels in
Figure~\ref{fig:events_io}.
We first observe that, as long as the results shown in
Figure~\ref{fig:events_io} is concerned, for relatively
small ($\lsim 0.1$) values of couplings,
the lightest neutrino mass dependence comes mainly from the
mass dependence in the visible part of the probabilities given in
Eqs.~(\ref{eq10}) and (\ref{parent-contribution-NO-IO}).
We remind the readers that
the daughter contributions coming from Eqs.~(\ref{eq5}) or ~(\ref{eq6})
in the effective probabilities shown
in Figure~\ref{fig:prob} is quite small, which should be
reflected in the event number distributions.

However, expressions shown in Eqs.~(\ref{eq10}) and
(\ref{parent-contribution-NO-IO}) are not useful to understand
the lightest neutrino mass dependence we can see in Figure~\ref{fig:events_io}
since the lifetime $\tau_i$
appears in these equations also depend on neutrino masses.
Therefore, we should take a closer look at the expressions of
$\Gamma$ functions given in the Appendix, taking into account that
$\Gamma_{ij} = m_i/(\tau_i E)$ for the decay mode of $\nu_i \to \nu_j$.

By looking into the expression of decay width $\Gamma$ functions
in Eq.~(\ref{eq:Gamma-W}),
we can say that the origin of the lightest neutrino mass dependence
comes from two parts: (i) the part which is given by the square of the mass of parent
neutrino, $m_i^2$, a factor common for both helicity flipping and
conserving processes, and (ii) the part which is described by
the dimensionless functions $f(x), h(x)$ and $k(x)$
shown in the Appendix, which have dependence on the both parent and daughter
masses of neutrinos as well as on the helicity of daughter neutrino,
if it is flipped or conserved.

Let us first take a look the part (ii) which looks more complicated.
For the case $g_{\rm S}~=~g_{\rm PS}$, the total rate coming from this
part is proportional to the sum of $(f(x)+h(x)+k(x))/x$ as we can see
from Eq.~(\ref{eq:Gamma-W}) in Appendix~\ref{apa}.
We observe that the variation of the lightest neutrino mass have little
impact on these functions (mainly due to $f(x)$ which is dominant),
and therefore induces little impact on the total rate,
at most a factor of $\sim$ 2-3 for both mass orderings (see Figure~1 in Ref.~\cite{Coloma:2017zpg}
for the NO where $f(x)/x, h(x)/x, k(x)/x$ is shown as a function of $x$).

On the other hand, the part (i), mass square of the parent neutrinos, ${m_i^2}$,
has stronger dependence on the lightest neutrino masses for the both mass orderings.
In the IO, for the case of $\nu_2 \to \nu_3$ decay, the parent mass is
$m_2=\sqrt{m_{\rm lightest}^2+\Delta m_{21}^2+|\Delta m_{31}^2|}$
which implies that the two different values of lightest neutrino
mass lead to $m^2_{2} \simeq \Delta m^2_{\rm atm} = 2.40 \times
10^{-3}$~eV$^2$ for $m_{ \text{lightest} } =10^{-3}$~eV, and $m^2_{2}
\simeq m_{\rm  lightest}^2 =10^{-2}$~eV$^2$ for $m_{ \text{lightest} } =10^{-1}$~eV. As a result, the decay width is an order of magnitude larger in the case
of $m_{ \text{lightest} } =10^{-1}$~eV, leading to the small but visible
difference between the left and right panels for the IO in
Figure~\ref{fig:events_io}.
In the NO, the situation is similar but $m_\text{lightest}$ dependence
is not visible in Figure~\ref{fig:events_no} because the impact of decay
itself is small due to small value of $\theta_{13}$ as explained in
the end of Section~\ref{Sec:what-new}.
We must remind the readers that these particular dependences of the decay rate on the parent neutrino mass may be model-dependent, which should be kept in mind in interpreting our results.

From Figure~\ref{fig:events_no}, for the case of NO, it can be seen that the KamLAND experiment is almost insensitive to the decay of $\overline{\nu}_3$ to $\overline{\nu}_2/\nu_2$ or $\overline{\nu}_1/\nu_1$. This is expected because the $\Delta m^2_{31}$-driven oscillations are averaged out at the baselines relevant to the KamLAND experiment, and therefore, the distortions in the spectrum due to decay cannot be seen.
Though a small pile-up of events due to daughter neutrinos is seen, KamLAND cannot place a useful bound on $\tau_{3}$ due to the appearance of daughter neutrinos in the NO case, because they are too small to be statistically significant.
For the case of the JUNO experiment, we find that for the NO there is a
very small effect of decay on the event rates. In the NO, small $s^2_{13}$ suppresses decay of $\overline{\nu}_3$ into $\overline{\nu}_2/\nu_2$ or $\overline{\nu}_1/\nu_{1}$, whose effect is mainly just to dampen the $\Delta m^2_{31}$-driven oscillations, as explored previously in Ref.~\cite{Abrahao:2015rba}.

In the case of IO, because of more than a factor of three longer average
baselines of KamLAND which leads to the larger effect of neutrino decay
than that for JUNO, KamLAND should be able to place a stronger
constraint on $\nu_{2}$ lifetime, if the number of events was similar to
that of JUNO. However, this advantage is largely compensated by much
lower statistics of KamLAND with
the total number of events, 2611 (including backgrounds)
\cite{Gando:2013nba}, which is smaller than those assumed for JUNO by a
factor of  54. 
Therefore, interpretation of the KamLAND bound on $\nu_{2}$ lifetime, which is only slightly better than JUNO as will be reported in Section~\ref{Sec:results}, must be done with care.
\begin{figure}[htbp!]
\centering
\includegraphics[scale=0.7]{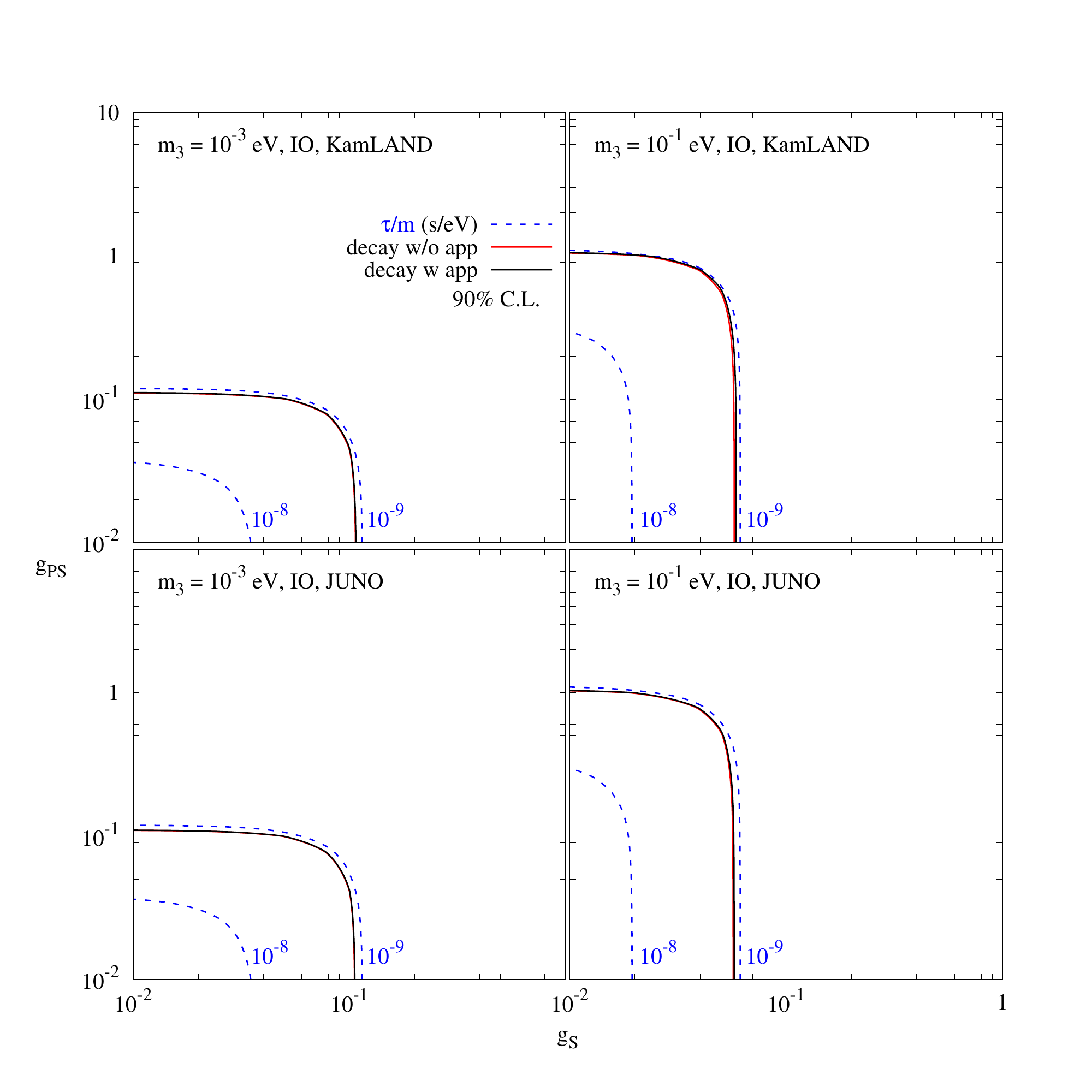}
\caption{\footnotesize{
Decay constraints and sensitivity plots for KamLAND (top panel) and JUNO (bottom panel). We show the $90\%$ C.L. contours in the $g_{\rm{S}}-g_{\rm{PS}}$ plane. The true mass ordering is assumed to be inverted. The left (right) panel is for $m_{\rm lightest}=10^{-3}$~eV ($10^{-1}$~eV). The red curve (labeled ``decay w/o app'') corresponds to visible decay including the contribution from the parent neutrinos only while the black curve (labeled ``decay w app'') corresponds to visible decay including the contributions from both parents as well as the daughter neutrinos. The blue curves show the respective values of $g_{\rm{S}},g_{\rm{PS}}$ which give $\left(\tau/m\right)_{\rm{heaviest}}= 10^{-8},10^{-9}$~s/eV.}
}
\label{fig:chisq_io}
\end{figure}

\section{Constraints on visible neutrino decay by KamLAND and JUNO}
\label{Sec:results}

In this section, we present the results of our analysis to obtain the
constraints on visible neutrino decay imposed by the KamLAND data, and
by a simulated data of JUNO assuming the total number of events equals
to 140,000 which can be obtained by the exposure of $\sim$ 220 GW$\cdot$years 
(total reactor thermal power times running period with $\sim$ 90\% of reactor avilability
assuming 100\% detection efficiency).
We exhibit the obtained constraints by drawing the $90\%$ C.L. exclusion contours in the $g_{\rm S}-g_{\rm PS}$ plane in Figure~\ref{fig:chisq_io} for IO, and in Figure~\ref{fig:chisq_no} for NO, for both KamLAND and JUNO. To translate these results to the constraints on the ratio of lifetime $\tau$ to the mass $m$, we show the equal $\tau/m$ contours as a function of $g_{\rm S}$ and $g_{\rm PS}$ in Figs.~\ref{fig:chisq_io} and \ref{fig:chisq_no}.

\subsection{Analysis procedure}
\label{Sec:procedure}

We now describe the numerical procedure for calculating the $\chi^2$ for excluding decay. For KamLAND, we consider the data presented in \cite{Gando:2013nba} while for JUNO, we simulate the ``true events rates'' assuming that neutrinos undergo only standard oscillations and that no neutrino decay occurs. To simulate the true events rates in the case of JUNO, we take the values of the oscillation parameters used in Section~\ref{eventsfeatures}.

We then ``fit'' the observed/simulated data with the calculated event rates assuming the existence of neutrino decay, by varying freely the values of $g_{\rm S}$ and $g_{\rm PS}$ in addition to varying the standard oscillation parameters $\theta_{12}$, $\theta_{13}$, $\Delta m^2_{21}$ and $\vert\Delta m^2_{31}\vert$ in their currently-allowed $3\sigma$ ranges. Note that in the fit we keep the test mass ordering same as the true one. These events rates are called the ``test event rates''. The binned-$\chi^2$ are calculated using GLoBES including the marginalization over the systematic uncertainties. We also add $\chi^2$ due to the Gaussian priors corresponding to the test oscillation parameters that are varied in the fit. The formula for the total $\chi^2$ is given by
\begin{eqnarray}
\chi^2_{\rm{total}} &=&
\sum_{i=1}^{n} 2\left[F_{i}\left(1+\xi_{1}+\xi_{2}\right)-
D_{i}+D_{i}\ln\left(D_{i}/F_{i}\left(1+\xi_{1}+\xi_{2}\right)\right)\right]
\nonumber \\
&+& \sum_{k}\left(\xi_k/\sigma_{k}\right)^2
+ \sum_{j}(\theta_{j} - \theta_{j}^{\rm{bf}})^2/\sigma_{j}^2.
\end{eqnarray}
Here, $n=17$ ($n=200$) is the total number of energy bins for KamLAND (JUNO). $F_{i}$ and
$D_{i}$ are the theoretical number of events and the observed number of
events, respectively, in a given $i$-th bin. $\xi_{k}$ are the
systematic uncertainty parameters with standard deviation $\sigma_{k}$;
and $\theta_{j}^{\rm{bf}}$ is the best fit value of a given oscillation
parameter $\theta_{j}$ (that are varied in the fit) with a $1\sigma$
uncertainty $\sigma_{j}$. For both KamLAND and JUNO, we consider an overall normalization
error of $\xi_1 = 5\%$ for signal and $20\%$ for the background events and an energy calibration error of $\xi_2 = 3\%$. For a given
choice of the test decay parameters $g_{\rm S}$ and $g_{\rm PS}$, we
select the least $\chi^2_{\rm{total}}$ that is obtained after
marginalizing over all the test oscillation parameters. The
$\Delta\chi^2$ for a given $g_{\rm S}$ and $g_{\rm PS}$ is obtained
through: $\Delta\chi^2=\chi^2_{\rm total}-\chi^2_{\rm
total,~smallest}$.
We show the resulting contours corresponding to
$\Delta\chi^2$ =4.61 for 2 degree of freedom (DOF)
as a function of the test $g_{\rm S}$ and $g_{\rm PS}$.

\subsection{KamLAND and JUNO bounds on neutrino decay: the inverted mass ordering}

\label{decay-at-juno}

We first discuss the potential of the experiments to exclude visible
decay for the case of IO, shown in Figure~\ref{fig:chisq_io}. The top
panels show the results for KamLAND while the bottom
panels show the results for JUNO. 
The left panels in these figures are for $m_{\rm{lightest}}=10^{-3}$~eV while the right panels are for $m_{\rm{lightest}}=10^{-1}$~eV.

From the top panels of Figure~\ref{fig:chisq_io}, we find that KamLAND
excludes $g_{\rm  S}~\gtrsim0.11$ and $g_{\rm PS}\gtrsim0.11$ for
$m_{\rm{lightest}}=10^{-3}$~eV at 90\% C.L. For
$m_{\rm{lightest}}=10^{-1}$~eV, $g_{\rm S}\gtrsim0.06$ and $g_{\rm
PS}\gtrsim1.00$ are excluded.
In either case, the constraints correspond
to the exclusion of $\tau/m~\lesssim1.1\times10^{-9}~\rm{s/eV}$. From
the lower panel, we see that for $m_{\rm{lightest}}=10^{-3}$~eV,
JUNO can exclude neutrino visible decay at $90\%$ C.L. for IO if $g_{\rm S}\gtrsim0.11$ and $g_{\rm PS}\gtrsim0.11$. For
$m_{\rm{lightest}}=10^{-1}$~eV,  if $g_{\rm
S}\gtrsim0.05$ and $g_{\rm PS}~\gtrsim1.00$,  JUNO can exclude
neutrino decay at $90~\%$~C.L..
For both KamLAND and JUNO, the constraint for the pseudo-scalar coupling
for $m_{\rm{lightest}}=10^{-1}$~eV is much weaker
because the functions $h(x)$ and $k(x)$ in $\Gamma$ are much
smaller than $f(x)$ for $m_{\rm{lightest}}=10^{-1}$~eV,
see Figure~1 in \cite{Coloma:2017zpg}.
Expressed in terms of
$\tau/m$, for IO, JUNO excludes
$\tau/m\lesssim1.1\times10^{-9}~\rm{s/eV}$,
which happened to be the same value as that of KamLAND.
We see that inclusion of the daughter neutrino contributions does not affect in any significant way the decay exclusion sensitivity. In all the panels, the two curves with and without the daughter's contributions are nearly coincident\footnote{This is not true in general. In the case of electron disappearance channel, the contribution of the neutrino decay to daughter neutrinos is suppressed in both NO (when the $\overline{\nu}_{3}$ state decays to $\overline{\nu}_{1}$ or $\overline{\nu}_{2}$) as well as IO (when $\overline{\nu}_{1}$ or $\overline{\nu}_{2}$ decays to $\overline{\nu}_{3}$) because the production as well as detection involves $\overline{\nu}_{e}$ which has a very small $\overline{\nu}_{3}$ content due to the smallness of $\vert U_{e3}\vert$. It was shown in Ref.~\cite{Coloma:2017zpg} that there can be significant daughter neutrinos in the $\nu_{\mu} \rightarrow \nu_{e}$ channel for the NO as the decay does not involve $\vert U_{e3}\vert$.}. 

It is remarkable that despite that
the number of events obtained by KamLAND used for our analysis is 54 times
smaller than that for JUNO (2611 vs 140,000),
both experiments give very similar bounds (sensitivities).
As mentioned at the end of the previous section, we understand that
this is mainly because of more than 3 times larger average baseline of KamLAND ($\sim$ 180 km)
compared to JUNO ($\sim$ 53 km) which can largely compensate
the much smaller statistics of KamLAND.

Let us now try to understand qualitatively the dependence of the value of $m_\text{lightest}$ on the sensitivities to $g_{\rm S}$ and $g_{\rm PS}$ we can see in Figure~\ref{fig:chisq_io}. Since the contributions from daughter neutrinos are small, we just need to pay attention to the decay width $\Gamma$ in Eq.~(\ref{eq:Gamma-W}), in particular for the helicity conserving case $r=s$ which is dominant.

\begin{figure}[htbp!]
\centering
\includegraphics[scale=0.7]{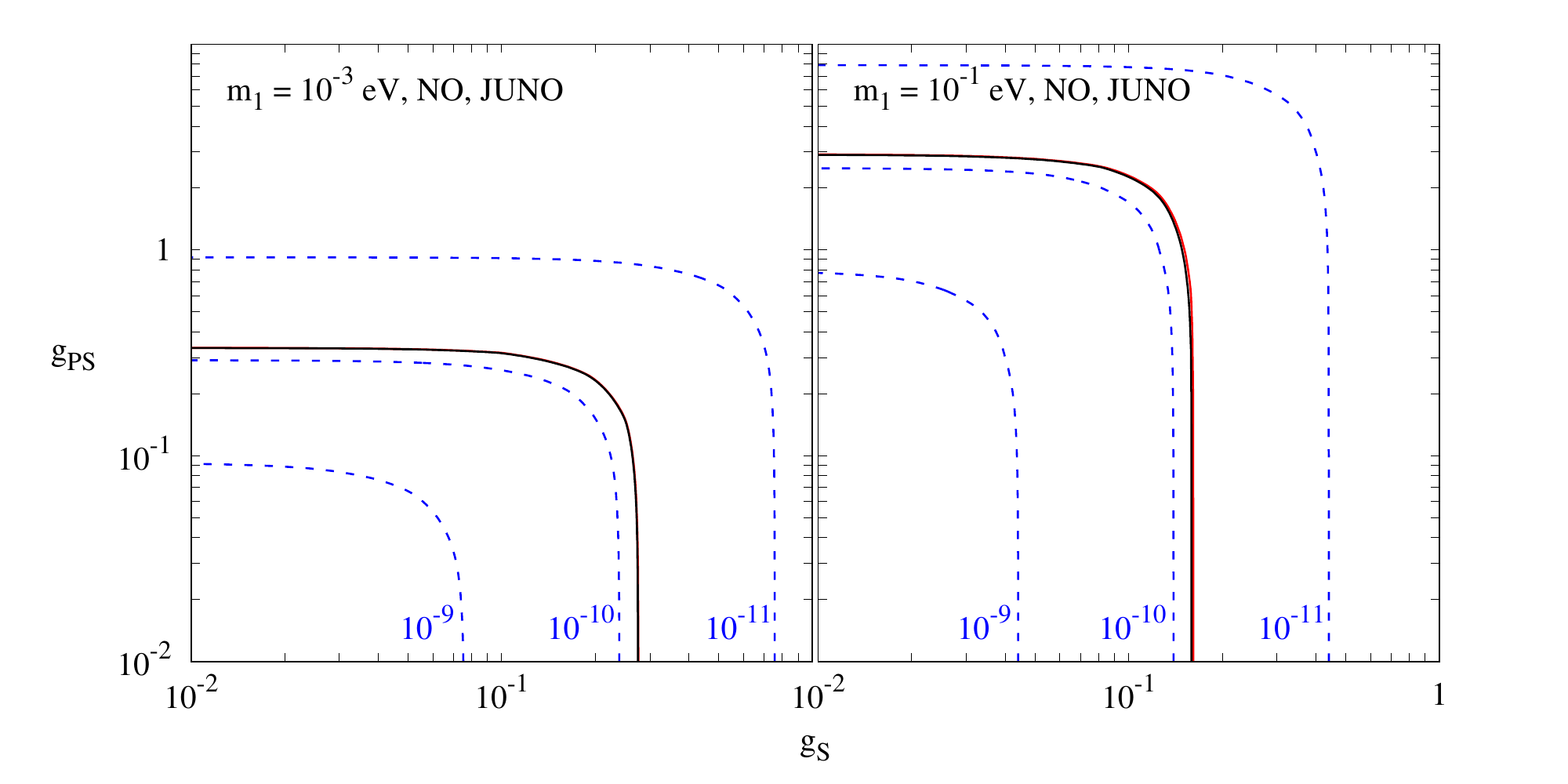}
\vspace{-6mm}
\caption{\footnotesize{Decay sensitivity plots for JUNO.
We show the $90\%$ C.L. contours in the $g_{\rm{S}}-g_{\rm{PS}}$ plane
for $m_{\rm lightest}=10^{-3}$~eV ($10^{-1}$~eV) in
the left (right) panel.
The true mass ordering is assumed to be normal.
The same notation apply as in Figure~\ref{fig:chisq_io}. The blue curves show the respective values of $g_{\rm{S}},g_{\rm{PS}}$ which give $\left(\tau/m\right)_{\rm{heaviest}}= 10^{-9},10^{-10},10^{-11}$~s/eV.}}
\label{fig:chisq_no}
\end{figure}
We first note that in the limit of vanishing $m_\text{lightest}$, which corresponds to $x\to \infty$, the functions $f(x)$, $h(x)$ and $k(x)$
given in eq.(\ref{eq:fgk}) tend to become equal. See also Figure~1 of Ref. \cite{Coloma:2017zpg}.
It implies that both couplings, $g_{\rm S}$ and $g_{\rm PS}$, contribute
equally to neutrino decay, which explains why the bounds are nearly
symmetric to $g_{\rm S}$ and $g_{\rm PS}$ in the case of
$m_\text{lightest}=10^{-3}$ eV.

On the other hand, as the assumed true value of $m_\text{lightest}$
is increased (or $x$ is decreased), the function $f(x)/x$ ($h(x)/x$) is
increased (decreased),
as can be seen from the first equation in (\ref{eq:Gamma-W})
and also from Figure~1 of Ref. \cite{Coloma:2017zpg}.
It means that the scalar (pseudo-scalar) coupling
$g_{\rm S}$ ($g_{\rm PS}$) becomes more (less) important for decay.
This is the reason why the bound on the scalar (pseudo-scalar)
coupling become tighter (milder) at a large value of
$m_\text{lightest}~=~10^{-1}$~eV independent of the mass orderings.

\subsection{JUNO bound on neutrino decay: the normal mass ordering }
\label{cons-kamland}

We consider only the JUNO experiment as it was shown previously (see Section~\ref{eventsfeatures}) that there is little sensitivity to the decay $\nu_{3}\rightarrow\nu_{1,2}$ in KamLAND.\footnote{
This point is reassured in the same numerical analysis as the case of IO.
}
JUNO will be able to place a limit on the decay effect
because of the much larger statistics and
good energy resolution which allows to detect
the damping-like effect in the $\Delta m^2_{31}$ driven oscillation
we can see in the left middle panel of Figure~\ref{fig:prob}.
From Figure \ref{fig:chisq_no}, we see that for
$m_{\rm{lightest}}=10^{-3}$~eV, JUNO can exclude neutrino visible decay
at $90\%$ C.L. if $g_{\rm S}\gtrsim0.28$ and $g_{\rm
PS}\gtrsim0.33$. For $m_{\rm{lightest}}=10^{-1}$~eV, JUNO can exclude
neutrino decay at $90\%$ C.L. if the mass ordering is normal and $g_{\rm
S}\gtrsim0.16$ and $g_{\rm PS}\gtrsim2.90$.
The constraint on $g_{\rm PS}$ is much weaker mainly due to the same
reason described in Section~\ref{cons-kamland} for the IO case.
Expressed in terms of $\tau/m$, we find that for NO, JUNO excludes $\tau/m\lesssim7.5\times10^{-11}~\rm{s/eV}$.\footnote{
This number is identical to the one obtained as the 95\% CL bound in ref.~\cite{Abrahao:2015rba}, but for invisible decay and for 1 DOF.
}
A comparison between Figs.~\ref{fig:chisq_io} and \ref{fig:chisq_no} indicates that the bounds crucially depend on the choice of the mass ordering: the constraints are milder in the NO by an order of magnitude. In the NO, as in the case of IO, the daughter neutrinos give essentially no contribution to exclusion of decay, leading to almost complete degeneracy of the red and the black curves in Figure~\ref{fig:chisq_no}.
\begin{table}[H]
\centering
\begin{tabular}{ l c c c l }
\hline
\hline
Experiment & (ordering, $m_{\rm{lightest}}$) & $g_{\rm{S}}$ & $g_{\rm{PS}}$ & $\tau/m$  (s/eV) \\
\hline
\hline
KamLAND & (IO, $10^{-3}$~eV) & $0.11$ & $0.11$ & $ 1.1\times 10^{-9}$  \\
KamLAND & (IO, $10^{-1}$~eV) & $0.06$ & $1.00$ & $1.1\times 10^{-9}$  \\
\hline
\hline
JUNO & (IO, $10^{-3}$~eV) & $0.11$ & $0.11$ & $1.1\times 10^{-9}$  \\
JUNO & (IO, $10^{-1}$~eV) & $0.05$ & $1.00$ & $1.1\times 10^{-9}$  \\
\hline
\hline
JUNO & (NO, $10^{-3}$~eV) & $0.28$ & $0.33$ & $7.5\times 10^{-11}$  \\
JUNO & (NO, $10^{-1}$~eV) & $0.16$ & $2.90$ & $7.5\times 10^{-11}$  \\
\hline
\hline
\end{tabular}
\caption{\label{tab:results-2DOF}
\footnotesize{
The $90\%$ C.L. upper bound on the couplings and the lower bound on the lifetime of active neutrinos for the given mass ordering and the values of
the lightest neutrino mass,
obtained through the sensitivity analyses of the
experiments KamLAND and JUNO.
The values of $g_{\rm{S}}$ and $g_{\rm{PS}}$
shown correspond to $\Delta\chi^2=4.61$ for 2 DOF. The value of $\tau/m$
shown is calculated for the given $m_{\rm{lightest}}$ and the obtained
value of $g_{\rm{S}}$ again for 2 DOF.
}}
\end{table}

In Table~\ref{tab:results-2DOF}, we summarize the constraints obtained
by KamLAND and JUNO  discussed in this section.

\section{Conclusions}
\label{Sec:summary}

In this work, we have discussed the effect of visible neutrino decay which can be detected by observing the positron energy spectrum due to the IBD reaction in reactor neutrino experiments. Modifications of the spectrum not only in the shape but also in the normalization are important. We have obtained the constraints on the lifetime of higher-mass state neutrinos ($\nu_{3}$ in the NO, and $\nu_{2}$ or $\nu_{1}$ in the IO) in medium and long-baseline reactor experiments.

We have used the Majoron model to calculate the neutrino decay rate.
We took the two experimental settings, KamLAND and
JUNO. They are the most relevant ones because of the long baseline,
$\sim180$ km for KamLAND, and high energy resolution $\lsim 3\%$
expected for JUNO in construction. For KamLAND we use the latest data,
and for JUNO we assume the exposure of about 220 GW$\cdot$years
which would produce $1.4\times10^{5}$ events.
In comparison with the constraints on neutrino decay often expressed
as the bound on $\tau/m$ ($\tau_{3} / m_{3}$ for the NO, and $\tau_{2} / m_{2}$ for the IO) at 90\% C.L. for 1 degree of freedom, we have provided the corresponding information in Table~\ref{tab:results-1DOF}.

\begin{table}[H]
\centering
\begin{tabular}{ l c l }
\hline
\hline
Experiment & (ordering, $m_{\rm{lightest}}$) & $\tau/m$  (s/eV) \\
\hline
\hline
KamLAND & (IO, $10^{-3}$~eV) & $1.4\times 10^{-9}$  \\
KamLAND & (IO, $10^{-1}$~eV) & $1.4\times 10^{-9}$  \\
\hline
\hline
JUNO & (IO, $10^{-3}$~eV) & $1.4\times 10^{-9}$  \\
JUNO & (IO, $10^{-1}$~eV) & $1.4\times 10^{-9}$  \\
\hline
\hline
JUNO & (NO, $10^{-3}$~eV) & $1.0\times 10^{-10}$  \\
JUNO & (NO, $10^{-1}$~eV) & $1.0\times 10^{-10}$  \\
\hline
\hline
\end{tabular}
\caption{\label{tab:results-1DOF}
\footnotesize{
The $90\%$ C.L. lower bound on the lifetime / mass
of active neutrinos for the given mass ordering and the values of the lightest neutrino mass, obtained through
the sensitivity analyses of the experiments KamLAND and JUNO.
The values of $\tau/m$ = $\tau_{3} / m_{3}$ ($\tau_{2} / m_{2}$)
for the normal (inverted) ordering, calculated for a given $m_{\rm{lightest}}$,}
shown correspond to $\Delta\chi^2=2.71$ for 1 DOF.
}
\end{table}

We found that the lifetime bounds depend crucially on whether the
neutrino mass ordering is normal or inverted.
Roughly speaking, the results we obtained for
JUNO shows that for the IO, the bounds are better than the one for the NO approximately by a factor of 20.
In looking into closer detail, KamLAND is insensitive to the decay of $\nu_{3}$ in the case of NO because of insufficient energy resolution to measure small wiggles of the atmospheric-scale high-frequency oscillations and statistically-insignificant pile-up of events due to daughter neutrinos at lower energies.
JUNO, on the other hand, can rule out $\tau/m \lesssim 1.0 \times
10^{-10}$ s/eV  for the $\nu_{3}$ mass eigenstate.
We note that this value is quite similar and consistent with the
bound of 9.3 $\times 10^{-11}$ s/eV for the same confidence level (90\% C.L.),
obtained in \cite{Abrahao:2015rba} where the invisible neutrino decay for JUNO was studied.
For the case of IO, the bounds of roughly the same order of
magnitude are obtained on the decay of the
$\nu_{2}$ mass eigenstate by both KamLAND and JUNO. 
We find that $\tau/m \lesssim 1.4 \times 10^{-9}$~s/eV is ruled out by 
both KamLAND and JUNO for the $\nu_{2}$ mass eigenstate.

We have observed that, in each of these cases, there is no significant improvement in the sensitivity by including the daughter neutrino contributions in the analyses.  It is because the effect is suppressed through $\vert U_{e3}\vert^2$,  which is small. However, in the case of IO, a decrease of $\overline{\nu}_{e}$ flux by the decay of $\overline{\nu}_{2}$ and $\overline{\nu}_{1}$ mass eigenstates produces a clear signature of neutrino decay, yielding a stringent lifetime bound mentioned above.
For the NO, neutrino decay acts merely as a damping effect of the fast $\Delta m^2_{31}$-driven oscillations both in JUNO and KamLAND, rendering detection of decay effect  harder, in particular, for the latter.
We also mention that our lifetime bound depends on $m_{\rm{lightest}}$ through $m_{\rm{lightest}}$ dependence of the decay rate.

\begin{acknowledgments}

SP, YPPS, and OLGP are thankful for the support of FAPESP
funding Grant  No. 2014/19164-6. OLGP was supported by FAPESP funding Grant 2016/08308-2, FAEPEX funding grant 2391/2017 and 2541/2019,  CNPq grants 304715/2016-6 and 306565/2019-6. YPPS is thankful for the support of FAPESP funding Grant No. 2017/05515-0 and 2019/22961-9.  SP is thankful for the support of FAPESP funding Grant No. 2017/02361-1. This
study was financed in part by the Coordena\c{c}\~ao de
Aperfei\c{c}oamento de Pessoal de N\'{\i}vel Superior - Brasil (CAPES) - Finance Code 001.

During this work, HM had been visiting Instituto de F{\'i}sica Gleb Wataghin, Universidade Estadual de Campinas in Brazil, Departamento de F\'{\i}sica, Pontif{\'\i}cia Universidade Cat{\'o}lica do Rio de Janeiro in Brazil, Instituto F\'{\i}sica Te\'{o}rica, UAM/CSIC in Madrid, Research Center for Cosmic Neutrinos, Institute for Cosmic Ray Research, University of Tokyo, before reaching Center for Neutrino Physics, Virginia Tech. He expresses deep gratitude to all of them for their hospitalities and supports. HN thanks the hospitality of the Fermilab Theoretical Department where the final part of this work was done.
\end{acknowledgments}
\appendix

\section{Auxiliary formulae}
\label{apa}
We have the auxiliary functions, respectively the decay rate, $\Gamma^{rs}_{ij}$ of $\nu^{(r)}_i \to \nu^{(r)}_j$ of initial mass state of mass $m_i$ and helicity {\it r} and final neutrino mass $m_j$ and helicity {\it s}  ,  and normalized spectrum of daughter distribution, $W^{rs}_{ij}(E_\alpha,E_\beta)$ with $E_{\alpha}$ the energy of initial state and $E_{\beta}$ the energy of the final neutrino~\cite{Lindner:2001fx,
PalomaresRuiz:2005vf,Coloma:2017zpg}

\begin{eqnarray}
\Gamma^{rs}_{ij}=
\left\{
\begin{array}{l l}
\frac{m_{i}^2}{16\pi E_{i}}
\bigg[(g_{\rm S}^{ij})^{2}\left(\frac{f(x_{ij})}{x_{ij}}\right)
+ (g_{\rm PS}^{ij})^{2}\left(\frac{h(x_{ij})}{x_{ij}}\right)
\bigg]
\quad \phantom{lala}& r=s , \\[0.5cm]
\frac{m_{i}^2}{16\pi E_{i}}
\left[
((g_{\rm S}^{ij})^{2}+(g_{\rm PS}^{ij})^{2})\left(\frac{k(x_{ij})}{x_{ij}}\right)
\right]  &  r \neq s
\\
\end{array}
\right.
\nonumber\\
W^{rs}_{ij}(E_\alpha,E_\beta) \equiv \dfrac{1}{\Gamma^{rs}_{ij}} \dfrac{d\Gamma^{rs}_{ij} (E_\alpha,E_\beta)}{dE_\beta}=
\left\{
\begin{array}{l l}
\left(\dfrac{1}{E_\alpha}\right)
\dfrac{ (g_{\rm S}^{ij})^2 (R + 2) + (g_{\rm PS}^{ij})^2 (R - 2) }{ (g_{\rm S}^{ij})^2 f(x_{ij}) + (g_{\rm PS}^{ij})^2  h(x_{ij}) } \quad \phantom{lala}& r=s , \\[0.5cm]
\left(\dfrac{1}{E_\alpha}\right)  \dfrac{  \frac{1}{x_{ij}} + x_{ij} - R
}{ k(x_{ij}) } &  r \neq s
\label{eq:Gamma-W}
\end{array}
\right.
\end{eqnarray}
where
\[
  R \equiv \dfrac{1}{x_{ij}}\dfrac{E_\alpha}{E_\beta} + x_{ij} \frac{E_\beta}{E_\alpha}
\]
with $x_{ij} \equiv m_i/m_j > 1$, and the functions $f(x), h(x),k(x)$ are defined in Eq.~\eqref{eq:fgk}.

\begin{eqnarray}
f(x) & = & \dfrac{x}{2}+2+\dfrac{2 \ln(x)}{x}-\dfrac{2}{x^2}-\dfrac{1}{(2x^3)} \, , \nonumber \\
h(x) & = & \dfrac{x}{2}-2+\dfrac{2 \ln(x)}{x}+\dfrac{2}{x^2}-\dfrac{1}{(2x^3)} \, , \label{eq:fgk}  \\
k(x) & = &\dfrac{x}{2}-\dfrac{2\ln(x)}{x}-\dfrac{1}{(2x^3)} \,. \nonumber
\end{eqnarray}

\bibliographystyle{JHEP}
\bibliography{references.bib}

\providecommand{\href}[2]{#2}\begingroup\raggedright\begin{thebibliography}{10}

\bibitem{Petcov:1976ff}
S.~T. Petcov, \emph{{The Processes $\mu \to e \gamma, \mu \to e e \bar{e},
  \nu^{\prime}\to \nu\gamma$ in the Weinberg-Salam Model with Neutrino
  Mixing}}, {\emph{Sov. J. Nucl. Phys.} {\bfseries 25} (1977) 340}.

\bibitem{Marciano:1977wx}
W.~J. Marciano and A.~I. Sanda, \emph{{Exotic Decays of the Muon and Heavy
  Leptons in Gauge Theories}},
  \href{http://dx.doi.org/10.1016/0370-2693(77)90377-X}{\emph{Phys. Lett.}
  {\bfseries 67B} (1977) 303--305}.

\bibitem{PhysRevD.16.1444}
B.~W. Lee and R.~E. Shrock, \emph{Natural suppression of symmetry violation in
  gauge theories: Muon- and electron-lepton-number nonconservation},
  \href{http://dx.doi.org/10.1103/PhysRevD.16.1444}{\emph{Phys. Rev. D}
  {\bfseries 16} (Sep, 1977) 1444--1473}.

\bibitem{Shrock:1982sc}
R.~E. Shrock, \emph{{Electromagnetic Properties and Decays of Dirac and
  Majorana Neutrinos in a General Class of Gauge Theories}},
  \href{http://dx.doi.org/10.1016/0550-3213(82)90273-5}{\emph{Nucl. Phys.}
  {\bfseries B206} (1982) 359--379}.

\bibitem{Frieman:1987as}
J.~A. Frieman, H.~E. Haber and K.~Freese, \emph{{Neutrino Mixing, Decays and
  Supernova Sn1987a}},
  \href{http://dx.doi.org/10.1016/0370-2693(88)91120-3}{\emph{Phys. Lett.}
  {\bfseries B200} (1988) 115}.

\bibitem{Hirata:1987hu}
{\scshape Kamiokande-II} collaboration, K.~S. Hirata et~al., \emph{{Observation
  of a Neutrino Burst from the Supernova SN 1987a}},
  \href{http://dx.doi.org/10.1103/PhysRevLett.58.1490}{\emph{Phys. Rev. Lett.}
  {\bfseries 58} (1987) 1490--1493}.

\bibitem{Bionta:1987qt}
{\scshape IMB Collaboration} collaboration, R.~M. Bionta et~al.,
  \emph{{Observation of a Neutrino Burst in Coincidence with Supernova SN 1987a
  in the Large Magellanic Cloud}},
  \href{http://dx.doi.org/10.1103/PhysRevLett.58.1494}{\emph{Phys. Rev. Lett.}
  {\bfseries 58} (1987) 1494}.

\bibitem{Berezhiani:1989za}
Z.~G. Berezhiani and A.~Y. Smirnov, \emph{{Matter Induced Neutrino Decay and
  Supernova {SN1987A}}},
  \href{http://dx.doi.org/10.1016/0370-2693(89)90052-X}{\emph{Phys. Lett.}
  {\bfseries B220} (1989) 279--284}.

\bibitem{Kachelriess:2000qc}
M.~Kachelriess, R.~Tomas and J.~W.~F. Valle, \emph{{Supernova bounds on Majoron
  emitting decays of light neutrinos}},
  \href{http://dx.doi.org/10.1103/PhysRevD.62.023004}{\emph{Phys. Rev.}
  {\bfseries D62} (2000) 023004},
  [\href{https://arxiv.org/abs/arXiv:hep-ph/0001039}{{\ttfamily
  arXiv:hep-ph/0001039}}].

\bibitem{Tomas:2001dh}
R.~Tomas, H.~Pas and J.~W.~F. Valle, \emph{{Generalized bounds on Majoron -
  neutrino couplings}},
  \href{http://dx.doi.org/10.1103/PhysRevD.64.095005}{\emph{Phys. Rev.}
  {\bfseries D64} (2001) 095005},
  [\href{https://arxiv.org/abs/arXiv:hep-ph/0103017}{{\ttfamily
  arXiv:hep-ph/0103017}}].

\bibitem{Lindner:2001th}
M.~Lindner, T.~Ohlsson and W.~Winter, \emph{{Decays of supernova neutrinos}},
  \href{http://dx.doi.org/10.1016/S0550-3213(01)00603-4}{\emph{Nucl. Phys.}
  {\bfseries B622} (2002) 429--456},
  [\href{https://arxiv.org/abs/arXiv:astro-ph/0105309}{{\ttfamily
  arXiv:astro-ph/0105309}}].

\bibitem{Ando:2003ie}
S.~Ando, \emph{{Decaying neutrinos and implications from the supernova relic
  neutrino observation}},
  \href{http://dx.doi.org/10.1016/j.physletb.2003.07.009}{\emph{Phys. Lett.}
  {\bfseries B570} (2003) 11},
  [\href{https://arxiv.org/abs/arXiv:hep-ph/0307169}{{\ttfamily
  arXiv:hep-ph/0307169}}].

\bibitem{Ando:2004qe}
S.~Ando, \emph{{Appearance of neutronization peak and decaying supernova
  neutrinos}}, \href{http://dx.doi.org/10.1103/PhysRevD.70.033004}{\emph{Phys.
  Rev.} {\bfseries D70} (2004) 033004},
  [\href{https://arxiv.org/abs/arXiv:hep-ph/0405200}{{\ttfamily
  arXiv:hep-ph/0405200}}].

\bibitem{Fogli:2004gy}
G.~L. Fogli, E.~Lisi, A.~Mirizzi and D.~Montanino, \emph{{Three generation
  flavor transitions and decays of supernova relic neutrinos}},
  \href{http://dx.doi.org/10.1103/PhysRevD.70.013001}{\emph{Phys. Rev.}
  {\bfseries D70} (2004) 013001},
  [\href{https://arxiv.org/abs/arXiv:hep-ph/0401227}{{\ttfamily
  arXiv:hep-ph/0401227}}].

\bibitem{deGouvea:2019goq}
A.~de~Gouvêa, I.~Martinez-Soler and M.~Sen, \emph{{Impact of neutrino decays
  on the supernova neutronization-burst flux}},
  \href{http://dx.doi.org/10.1103/PhysRevD.101.043013}{\emph{Phys. Rev. D}
  {\bfseries 101} (2020) 043013},
  [\href{https://arxiv.org/abs/1910.01127}{{\ttfamily 1910.01127}}].

\bibitem{Bahcall:1972my}
J.~N. Bahcall, N.~Cabibbo and A.~Yahil, \emph{{Are neutrinos stable
  particles?}}, \href{http://dx.doi.org/10.1103/PhysRevLett.28.316}{\emph{Phys.
  Rev. Lett.} {\bfseries 28} (1972) 316--318}.

\bibitem{Raghavan:1987uh}
R.~S. Raghavan, X.-G. He and S.~Pakvasa, \emph{{MSW Catalyzed Neutrino Decay}},
  \href{http://dx.doi.org/10.1103/PhysRevD.38.1317}{\emph{Phys. Rev.}
  {\bfseries D38} (1988) 1317--1320}.

\bibitem{Berezhiani:1991vk}
Z.~G. Berezhiani, G.~Fiorentini, M.~Moretti and A.~Rossi, \emph{{Fast neutrino
  decay and solar neutrino detectors}},
  \href{http://dx.doi.org/10.1007/BF01559483}{\emph{Z. Phys.} {\bfseries C54}
  (1992) 581--586}.

\bibitem{Joshipura:1992vn}
A.~S. Joshipura and S.~D. Rindani, \emph{{Fast neutrino decay in the minimal
  seesaw model}}, \href{http://dx.doi.org/10.1103/PhysRevD.46.3000}{\emph{Phys.
  Rev.} {\bfseries D46} (1992) 3000--3007},
  [\href{https://arxiv.org/abs/hep-ph/9205220}{{\ttfamily hep-ph/9205220}}].

\bibitem{Acker:1992eh}
A.~Acker, A.~Joshipura and S.~Pakvasa, \emph{{A Neutrino decay model, solar
  anti-neutrinos and atmospheric neutrinos}},
  \href{http://dx.doi.org/10.1016/0370-2693(92)91520-J}{\emph{Phys. Lett.}
  {\bfseries B285} (1992) 371--375}.

\bibitem{Berezhiani:1992ry}
Z.~G. Berezhiani, G.~Fiorentini, A.~Rossi and M.~Moretti, \emph{{Neutrino decay
  solution of the solar neutrino problem revisited}}, {\emph{JETP Lett.}
  {\bfseries 55} (1992) 151--156}.

\bibitem{Berezhiani:1993iy}
Z.~G. Berezhiani and A.~Rossi, \emph{{Matter induced neutrino decay: New
  candidate for the solution to the solar neutrino problem}},  in \emph{{4th
  International Symposium on Neutrino Telescopes}}, pp.~123--135, 6, 1993.
\newblock \href{https://arxiv.org/abs/hep-ph/9306278}{{\ttfamily
  hep-ph/9306278}}.

\bibitem{Choubey:2000an}
S.~Choubey, S.~Goswami and D.~Majumdar, \emph{{Status of the neutrino decay
  solution to the solar neutrino problem}},
  \href{http://dx.doi.org/10.1016/S0370-2693(00)00608-0}{\emph{Phys. Lett.}
  {\bfseries B484} (2000) 73--78},
  [\href{https://arxiv.org/abs/arXiv:hep-ph/0004193}{{\ttfamily
  arXiv:hep-ph/0004193}}].

\bibitem{Bandyopadhyay:2001ct}
A.~Bandyopadhyay, S.~Choubey and S.~Goswami, \emph{{MSW mediated neutrino decay
  and the solar neutrino problem}},
  \href{http://dx.doi.org/10.1103/PhysRevD.63.113019}{\emph{Phys.Rev.}
  {\bfseries D63} (2001) 113019}.

\bibitem{Beacom:2002cb}
J.~F. Beacom and N.~F. Bell, \emph{{Do solar neutrinos decay?}},
  \href{http://dx.doi.org/10.1103/PhysRevD.65.113009}{\emph{Phys.Rev.}
  {\bfseries D65} (2002) 113009}.

\bibitem{Joshipura:2002fb}
A.~S. Joshipura, E.~Masso and S.~Mohanty, \emph{{Constraints on decay plus
  oscillation solutions of the solar neutrino problem}},
  \href{http://dx.doi.org/10.1103/PhysRevD.66.113008}{\emph{Phys.Rev.}
  {\bfseries D66} (2002) 113008}.

\bibitem{Bandyopadhyay:2002qg}
A.~Bandyopadhyay, S.~Choubey and S.~Goswami, \emph{{Neutrino decay confronts
  the SNO data}},
  \href{http://dx.doi.org/10.1016/S0370-2693(03)00044-3}{\emph{Phys.Lett.}
  {\bfseries B555} (2003) 33--42}.

\bibitem{Das:2010sd}
C.~R. Das and J.~Pulido, \emph{{Improving LMA predictions with non-standard
  interactions: Neutrino decay in solar matter?}},
  \href{http://dx.doi.org/10.1103/PhysRevD.83.053009}{\emph{Phys.Rev.}
  {\bfseries D83} (2011) 053009}.

\bibitem{Berryman:2014qha}
J.~M. Berryman, A.~de~Gouv\^ea and D.~Hernandez, \emph{{Solar Neutrinos and the
  Decaying Neutrino Hypothesis}},
  \href{http://dx.doi.org/10.1103/PhysRevD.92.073003}{\emph{Phys. Rev.}
  {\bfseries D92} (2015) 073003},
  [\href{https://arxiv.org/abs/1411.0308}{{\ttfamily 1411.0308}}].

\bibitem{Picoreti:2015ika}
R.~Picoreti, M.~M. Guzzo, P.~C. de~Holanda and O.~L.~G. Peres, \emph{{Neutrino
  Decay and Solar Neutrino Seasonal Effect}},
  \href{http://dx.doi.org/10.1016/j.physletb.2016.08.007}{\emph{Phys. Lett.}
  {\bfseries B761} (2016) 70--73},
  [\href{https://arxiv.org/abs/arXiv:1506.08158}{{\ttfamily
  arXiv:1506.08158}}].

\bibitem{Aharmim:2018fme}
{\scshape SNO} collaboration, B.~Aharmim et~al., \emph{{Constraints on Neutrino
  Lifetime from the Sudbury Neutrino Observatory}},
  \href{http://dx.doi.org/10.1103/PhysRevD.99.032013}{\emph{Phys. Rev.}
  {\bfseries D99} (2019) 032013},
  [\href{https://arxiv.org/abs/1812.01088}{{\ttfamily 1812.01088}}].

\bibitem{Funcke:2019grs}
L.~Funcke, G.~Raffelt and E.~Vitagliano, \emph{{Distinguishing Dirac and
  Majorana neutrinos by their decays via Nambu-Goldstone bosons in the
  gravitational-anomaly model of neutrino masses}},
  \href{http://dx.doi.org/10.1103/PhysRevD.101.015025}{\emph{Phys. Rev. D}
  {\bfseries 101} (2020) 015025},
  [\href{https://arxiv.org/abs/1905.01264}{{\ttfamily 1905.01264}}].

\bibitem{Hannestad:2005ex}
S.~Hannestad and G.~Raffelt, \emph{{Constraining invisible neutrino decays with
  the cosmic microwave background}},
  \href{http://dx.doi.org/10.1103/PhysRevD.72.103514}{\emph{Phys. Rev.}
  {\bfseries D72} (2005) 103514},
  [\href{https://arxiv.org/abs/hep-ph/0509278}{{\ttfamily hep-ph/0509278}}].

\bibitem{Baerwald:2012kc}
P.~Baerwald, M.~Bustamante and W.~Winter, \emph{{Neutrino Decays over
  Cosmological Distances and the Implications for Neutrino Telescopes}},
  \href{http://dx.doi.org/10.1088/1475-7516/2012/10/020}{\emph{JCAP} {\bfseries
  1210} (2012) 020}, [\href{https://arxiv.org/abs/arXiv:1208.4600}{{\ttfamily
  arXiv:1208.4600}}].

\bibitem{Dorame:2013lka}
L.~Dorame, O.~G. Miranda and J.~W.~F. Valle, \emph{{Invisible decays of
  ultra-high energy neutrinos}},
  \href{http://dx.doi.org/10.3389/fphy.2013.00025}{\emph{Front.in Phys.}
  {\bfseries 1} (2013) 25},
  [\href{https://arxiv.org/abs/arXiv:1303.4891}{{\ttfamily arXiv:1303.4891}}].

\bibitem{Bustamante:2016ciw}
M.~Bustamante, J.~F. Beacom and K.~Murase, \emph{{Testing decay of
  astrophysical neutrinos with incomplete information}},
  \href{http://dx.doi.org/10.1103/PhysRevD.95.063013}{\emph{Phys. Rev.}
  {\bfseries D95} (2017) 063013},
  [\href{https://arxiv.org/abs/1610.02096}{{\ttfamily 1610.02096}}].

\bibitem{Pagliaroli:2015rca}
G.~Pagliaroli, A.~Palladino, F.~L. Villante and F.~Vissani, \emph{{Testing
  nonradiative neutrino decay scenarios with IceCube data}},
  \href{http://dx.doi.org/10.1103/PhysRevD.92.113008}{\emph{Phys. Rev.}
  {\bfseries D92} (2015) 113008},
  [\href{https://arxiv.org/abs/arXiv:1506.02624}{{\ttfamily
  arXiv:1506.02624}}].

\bibitem{Escudero:2019gfk}
M.~Escudero and M.~Fairbairn, \emph{{Cosmological Constraints on Invisible
  Neutrino Decays Revisited}},
  \href{http://dx.doi.org/10.1103/PhysRevD.100.103531}{\emph{Phys. Rev.}
  {\bfseries D100} (2019) 103531},
  [\href{https://arxiv.org/abs/1907.05425}{{\ttfamily 1907.05425}}].

\bibitem{Barger:1998xk}
V.~D. Barger, J.~G. Learned, S.~Pakvasa and T.~J. Weiler, \emph{{Neutrino decay
  as an explanation of atmospheric neutrino observations}},
  \href{http://dx.doi.org/10.1103/PhysRevLett.82.2640}{\emph{Phys. Rev. Lett.}
  {\bfseries 82} (1999) 2640--2643},
  [\href{https://arxiv.org/abs/arXiv:astro-ph/9810121}{{\ttfamily
  arXiv:astro-ph/9810121}}].

\bibitem{Fogli:1999qt}
G.~L. Fogli, E.~Lisi, A.~Marrone and G.~Scioscia, \emph{{Super-Kamiokande data
  and atmospheric neutrino decay}},
  \href{http://dx.doi.org/10.1103/PhysRevD.59.117303}{\emph{Phys. Rev.}
  {\bfseries D59} (1999) 117303},
  [\href{https://arxiv.org/abs/arXiv:hep-ph/9902267}{{\ttfamily
  arXiv:hep-ph/9902267}}].

\bibitem{Meloni:2006gv}
D.~Meloni and T.~Ohlsson, \emph{{Neutrino flux ratios at neutrino telescopes:
  The Role of uncertainties of neutrino mixing parameters and applications to
  neutrino decay}},
  \href{http://dx.doi.org/10.1103/PhysRevD.75.125017}{\emph{Phys. Rev.}
  {\bfseries D75} (2007) 125017},
  [\href{https://arxiv.org/abs/arXiv:hep-ph/0612279}{{\ttfamily
  arXiv:hep-ph/0612279}}].

\bibitem{Maltoni:2008jr}
M.~Maltoni and W.~Winter, \emph{{Testing neutrino oscillations plus decay with
  neutrino telescopes}},
  \href{http://dx.doi.org/10.1088/1126-6708/2008/07/064}{\emph{JHEP} {\bfseries
  0807} (2008) 064}, [\href{https://arxiv.org/abs/arXiv:0803.2050}{{\ttfamily
  arXiv:0803.2050}}].

\bibitem{GonzalezGarcia:2008ru}
M.~C. Gonzalez-Garcia and M.~Maltoni, \emph{{Status of Oscillation plus Decay
  of Atmospheric and Long-Baseline Neutrinos}},
  \href{http://dx.doi.org/10.1016/j.physletb.2008.04.041}{\emph{Phys. Lett.}
  {\bfseries B663} (2008) 405--409},
  [\href{https://arxiv.org/abs/arXiv:0802.3699}{{\ttfamily arXiv:0802.3699}}].

\bibitem{Choubey:2017eyg}
S.~Choubey, S.~Goswami, C.~Gupta, S.~M. Lakshmi and T.~Thakore,
  \emph{{Sensitivity to neutrino decay with atmospheric neutrinos at the
  INO-ICAL detector}},
  \href{http://dx.doi.org/10.1103/PhysRevD.97.033005}{\emph{Phys. Rev.}
  {\bfseries D97} (2018) 033005},
  [\href{https://arxiv.org/abs/1709.10376}{{\ttfamily 1709.10376}}].

\bibitem{Denton:2018aml}
P.~B. Denton and I.~Tamborra, \emph{{Invisible Neutrino Decay Could Resolve
  IceCube’s Track and Cascade Tension}},
  \href{http://dx.doi.org/10.1103/PhysRevLett.121.121802}{\emph{Phys. Rev.
  Lett.} {\bfseries 121} (2018) 121802},
  [\href{https://arxiv.org/abs/1805.05950}{{\ttfamily 1805.05950}}].

\bibitem{Choubey:2018kah}
S.~Choubey, S.~Goswami, C.~Gupta, L.~S. Mohan and T.~Thakore, \emph{{Study of
  invisible neutrino decay and oscillation in the presence of matter with a 50
  kton magnetised iron detector}},
  \href{http://dx.doi.org/10.22323/1.295.0147}{\emph{PoS} {\bfseries
  NuFact2017} (2018) 147}.

\bibitem{Gomes:2014yua}
R.~A. Gomes, A.~L.~G. Gomes and O.~L.~G. Peres, \emph{{Constraints on neutrino
  decay lifetime using long-baseline charged and neutral current data}},
  \href{http://dx.doi.org/10.1016/j.physletb.2014.12.014}{\emph{Phys. Lett.}
  {\bfseries B740} (2015) 345--352},
  [\href{https://arxiv.org/abs/arXiv:1407.5640}{{\ttfamily arXiv:1407.5640}}].

\bibitem{Phdabner}
A.~L.~G. Gomes, ``{Master Thesis in portuguese, : Limites nos parâmetros do
  modelo de oscilação com decaimento de neutrinos usando os dados do
  experimento MINOS}.'' In portuguese,
  \url{https://repositorio.bc.ufg.br/tede/handle/tede/7509}, 2014.

\bibitem{Pagliaroli:2016zab}
G.~Pagliaroli, N.~Di~Marco and M.~Mannarelli, \emph{{Enhanced tau neutrino
  appearance through invisible decay}},
  \href{http://dx.doi.org/10.1103/PhysRevD.93.113011}{\emph{Phys. Rev.}
  {\bfseries D93} (2016) 113011},
  [\href{https://arxiv.org/abs/1603.08696}{{\ttfamily 1603.08696}}].

\bibitem{Gago:2017zzy}
A.~M. Gago, R.~A. Gomes, A.~L.~G. Gomes, J.~Jones-Perez and O.~L.~G. Peres,
  \emph{{Visible neutrino decay in the light of appearance and disappearance
  long baseline experiments}},
  \href{http://dx.doi.org/10.1007/JHEP11(2017)022}{\emph{JHEP} {\bfseries 11}
  (2017) 022}, [\href{https://arxiv.org/abs/1705.03074}{{\ttfamily
  1705.03074}}].

\bibitem{Choubey:2017dyu}
S.~Choubey, S.~Goswami and D.~Pramanik, \emph{{A study of invisible neutrino
  decay at DUNE and its effects on $\theta_{23}$ measurement}},
  \href{http://dx.doi.org/10.1007/JHEP02(2018)055}{\emph{JHEP} {\bfseries 02}
  (2018) 055}, [\href{https://arxiv.org/abs/1705.05820}{{\ttfamily
  1705.05820}}].

\bibitem{Choubey:2018cfz}
S.~Choubey, D.~Dutta and D.~Pramanik, \emph{{Invisible neutrino decay in the
  light of {\rm NOvA} and T2K data}},
  \href{http://dx.doi.org/10.1007/JHEP08(2018)141}{\emph{JHEP} {\bfseries 08}
  (2018) 141}, [\href{https://arxiv.org/abs/1805.01848}{{\ttfamily
  1805.01848}}].

\bibitem{deSalas:2018kri}
P.~F. de~Salas, S.~Pastor, C.~A. Ternes, T.~Thakore and M.~Tórtola,
  \emph{{Constraining the invisible neutrino decay with KM3NeT-ORCA}},
  \href{http://dx.doi.org/10.1016/j.physletb.2018.12.066}{\emph{Phys. Lett.}
  {\bfseries B789} (2019) 472--479},
  [\href{https://arxiv.org/abs/1810.10916}{{\ttfamily 1810.10916}}].

\bibitem{Tang:2018rer}
J.~Tang, T.-C. Wang and Y.~Zhang, \emph{{Invisible neutrino decays at the
  MOMENT experiment}},
  \href{http://dx.doi.org/10.1007/JHEP04(2019)004}{\emph{JHEP} {\bfseries 04}
  (2019) 004}, [\href{https://arxiv.org/abs/1811.05623}{{\ttfamily
  1811.05623}}].

\bibitem{Abrahao:2015rba}
T.~Abrah\~ao, H.~Minakata, H.~Nunokawa and A.~A. Quiroga, \emph{{Constraint on
  Neutrino Decay with Medium-Baseline Reactor Neutrino Oscillation
  Experiments}}, \href{http://dx.doi.org/10.1007/JHEP11(2015)001}{\emph{JHEP}
  {\bfseries 11} (2015) 001},
  [\href{https://arxiv.org/abs/1506.02314}{{\ttfamily 1506.02314}}].

\bibitem{Coloma:2017zpg}
P.~Coloma and O.~L.~G. Peres, \emph{{Visible neutrino decay at DUNE}},
  \href{https://arxiv.org/abs/1705.03599}{{\ttfamily 1705.03599}}.

\bibitem{Ascencio-Sosa:2018lbk}
M.~V. Ascencio-Sosa, A.~M. Calatayud-Cadenillas, A.~M. Gago and
  J.~Jones-P\'erez, \emph{{Matter effects in neutrino visible decay at future
  long-baseline experiments}},
  \href{http://dx.doi.org/10.1140/epjc/s10052-018-6276-0}{\emph{Eur. Phys. J.}
  {\bfseries C78} (2018) 809},
  [\href{https://arxiv.org/abs/1805.03279}{{\ttfamily 1805.03279}}].

\bibitem{Huang:2018nxj}
G.-Y. Huang and S.~Zhou, \emph{{Constraining Neutrino Lifetimes and Magnetic
  Moments via Solar Neutrinos in the Large Xenon Detectors}},
  \href{http://dx.doi.org/10.1088/1475-7516/2019/02/024}{\emph{JCAP} {\bfseries
  1902} (2019) 024}, [\href{https://arxiv.org/abs/1810.03877}{{\ttfamily
  1810.03877}}].

\bibitem{Chikashige:1980qk}
Y.~Chikashige, R.~N. Mohapatra and R.~D. Peccei, \emph{{Spontaneously Broken
  Lepton Number and Cosmological Constraints on the Neutrino Mass Spectrum}},
  \href{http://dx.doi.org/10.1103/PhysRevLett.45.1926}{\emph{Phys. Rev. Lett.}
  {\bfseries 45} (1980) 1926}.

\bibitem{Chikashige:1980ui}
Y.~Chikashige, R.~N. Mohapatra and R.~D. Peccei, \emph{{Are There Real
  Goldstone Bosons Associated with Broken Lepton Number?}},
  \href{http://dx.doi.org/10.1016/0370-2693(81)90011-3}{\emph{Phys. Lett.}
  {\bfseries B98} (1981) 265--268}.

\bibitem{Schechter:1981cv}
J.~Schechter and J.~W.~F. Valle, \emph{{Neutrino Decay and Spontaneous
  Violation of Lepton Number}},
  \href{http://dx.doi.org/10.1103/PhysRevD.25.774}{\emph{Phys. Rev.} {\bfseries
  D25} (1982) 774}.

\bibitem{Gelmini:1980re}
G.~B. Gelmini and M.~Roncadelli, \emph{{Left-Handed Neutrino Mass Scale and
  Spontaneously Broken Lepton Number}},
  \href{http://dx.doi.org/10.1016/0370-2693(81)90559-1}{\emph{Phys. Lett.}
  {\bfseries B99} (1981) 411--415}.

\bibitem{Gelmini:1983ea}
G.~B. Gelmini and J.~W.~F. Valle, \emph{{Fast Invisible Neutrino Decays}},
  \href{http://dx.doi.org/10.1016/0370-2693(84)91258-9}{\emph{Phys. Lett.}
  {\bfseries B142} (1984) 181}.

\bibitem{1988SvA....32..127D}
A.~G. {Doroshkevich}, A.~A. {Klypin} and M.~Y. {Khlopov}, \emph{{Cosmological
  Models with Unstable Neutrinos}}, {\emph{Sov. Astron.} {\bfseries 32} (Apr.,
  1988) 127}.

\bibitem{Berezhiani:1990sy}
Z.~G. Berezhiani and M.~{\relax Yu}. Khlopov, \emph{{Physics of cosmological
  dark matter in the theory of broken family symmetry. (In Russian)}},
  {\emph{Sov. J. Nucl. Phys.} {\bfseries 52} (1990) 60--64}.

\bibitem{Dias:2005jm}
A.~G. Dias, A.~Doff, C.~A. de~S.~Pires and P.~S. Rodrigues~da Silva,
  \emph{{Neutrino decay and neutrinoless double beta decay in a 3-3-1 model}},
  \href{http://dx.doi.org/10.1103/PhysRevD.72.035006}{\emph{Phys. Rev.}
  {\bfseries D72} (2005) 035006},
  [\href{https://arxiv.org/abs/hep-ph/0503014}{{\ttfamily hep-ph/0503014}}].

\bibitem{Gando:2013nba}
{\scshape KamLAND Collaboration} collaboration, A.~Gando et~al., \emph{{Reactor
  On-Off Antineutrino Measurement with KamLAND}},
  \href{http://dx.doi.org/10.1103/PhysRevD.88.033001}{\emph{Phys. Rev.}
  {\bfseries D88} (2013) 033001},
  [\href{https://arxiv.org/abs/arXiv:1303.4667}{{\ttfamily arXiv:1303.4667}}].

\bibitem{An:2015jdp}
{\scshape JUNO} collaboration, F.~An et~al., \emph{{Neutrino Physics with
  JUNO}}, \href{http://dx.doi.org/10.1088/0954-3899/43/3/030401}{\emph{J.
  Phys.} {\bfseries G43} (2016) 030401},
  [\href{https://arxiv.org/abs/1507.05613}{{\ttfamily 1507.05613}}].

\bibitem{Lindner:2001fx}
M.~Lindner, T.~Ohlsson and W.~Winter, \emph{{A Combined treatment of neutrino
  decay and neutrino oscillations}},
  \href{http://dx.doi.org/10.1016/S0550-3213(01)00237-1}{\emph{Nucl. Phys.}
  {\bfseries B607} (2001) 326--354},
  [\href{https://arxiv.org/abs/hep-ph/0103170}{{\ttfamily hep-ph/0103170}}].

\bibitem{Kim:1990km}
C.~W. Kim and W.~P. Lam, \emph{{Some remarks on neutrino decay via a
  Nambu-Goldstone boson}},
  \href{http://dx.doi.org/10.1142/S0217732390000354}{\emph{Mod. Phys. Lett.}
  {\bfseries A5} (1990) 297--299}.

\bibitem{PalomaresRuiz:2005vf}
S.~Palomares-Ruiz, S.~Pascoli and T.~Schwetz, \emph{{Explaining LSND by a
  decaying sterile neutrino}},
  \href{http://dx.doi.org/10.1088/1126-6708/2005/09/048}{\emph{JHEP} {\bfseries
  0509} (2005) 048},
  [\href{https://arxiv.org/abs/arXiv:hep-ph/0505216}{{\ttfamily
  arXiv:hep-ph/0505216}}].

\bibitem{Tanabashi:2018oca}
{\scshape Particle Data Group} collaboration, M.~Tanabashi et~al.,
  \emph{{Review of Particle Physics}},
  \href{http://dx.doi.org/10.1103/PhysRevD.98.030001}{\emph{Phys. Rev.}
  {\bfseries D98} (2018) 030001}.

\bibitem{Huber:2004ka}
P.~Huber, M.~Lindner and W.~Winter, \emph{{Simulation of long-baseline neutrino
  oscillation experiments with GLoBES (General Long Baseline Experiment
  Simulator)}},
  \href{http://dx.doi.org/10.1016/j.cpc.2005.01.003}{\emph{Comput. Phys.
  Commun.} {\bfseries 167} (2005) 195},
  [\href{https://arxiv.org/abs/hep-ph/0407333}{{\ttfamily hep-ph/0407333}}].

\bibitem{Huber:2007ji}
P.~Huber, J.~Kopp, M.~Lindner, M.~Rolinec and W.~Winter, \emph{{New features in
  the simulation of neutrino oscillation experiments with GLoBES 3.0: General
  Long Baseline Experiment Simulator}},
  \href{http://dx.doi.org/10.1016/j.cpc.2007.05.004}{\emph{Comput. Phys.
  Commun.} {\bfseries 177} (2007) 432--438},
  [\href{https://arxiv.org/abs/hep-ph/0701187}{{\ttfamily hep-ph/0701187}}].

\bibitem{deSalas:2017kay}
P.~F. de~Salas, D.~V. Forero, C.~A. Ternes, M.~Tortola and J.~W.~F. Valle,
  \emph{{Status of neutrino oscillations 2018: 3$\sigma$ hint for normal mass
  ordering and improved CP sensitivity}},
  \href{http://dx.doi.org/10.1016/j.physletb.2018.06.019}{\emph{Phys. Lett.}
  {\bfseries B782} (2018) 633--640},
  [\href{https://arxiv.org/abs/1708.01186}{{\ttfamily 1708.01186}}].

\end{thebibliography}\endgroup

\end{document}